%% file: EXO-24-028_temp.tex
\pdfoutput=1
\documentclass[11pt,twoside,a4paper,cmspaper,final,collab]{cms-tdr}
\def\svnVersion{2d021b8}\def\svnDate{2026/08/24}\def\cmsCernNoTag{CERN-EP-2026-048}\def\cmsCernDate{\today}\def\cmsMessage{Published in the Journal of High Energy Physics as \href{http://dx.doi.org/10.1007/JHEP08(2026)098}{\doi{10.1007/JHEP08(2026)098}.}}
\begin{document}\cmsNoteHeader{EXO-24-028}

\newlength\cmsFigWidth
\setlength\cmsFigWidth{0.4\textwidth}

\newcommand{\st}{\ensuremath{S_\text{T}}\xspace}
\newcommand{\stmin}{\ensuremath{S^\text{min}_\text{T}}\xspace}
\newcommand{\md}{\ensuremath{M_\text{D}}\xspace}
\newcommand{\mbh}{\ensuremath{M_\text{BH}}\xspace}
\newcommand{\ncs}{\ensuremath{N_\text{CS}}\xspace}
\newcommand{\pncs}{\ensuremath{p({N_\text{CS}})}\xspace}
\newcommand{\NFAIL}{\ensuremath{N_{\text{FAIL}}^{\text{bkg}}(j)}\xspace}
\newcommand{\NPASS}{\ensuremath{N_{\text{PASS}}^{\text{bkg}}(j)}\xspace}
\newcommand{\RPF}{\ensuremath{R_{\text{P/F}}}\xspace}
\newcommand{\RPFj}{\ensuremath{R_{\text{P/F}}(j)}\xspace}
\newcommand{\mbhmin}{\ensuremath{M_{\text{BH}}^{\text{min}}}\xspace}
\newcommand{\ms}{\ensuremath{M_{\text{S}}}\xspace}
\newcommand{\gs}{\ensuremath{g_{\text{S}}}\xspace}
\newcommand{\psph}{\ensuremath{p_{\text{sph}}}\xspace}
\newcommand{\esph}{\ensuremath{E_{\text{sph}}}\xspace}
\newcommand{\nmax}{\ensuremath{n^{\text{max}}}\xspace}

\cmsNoteHeader{EXO-24-028}
\title{A search for microscopic black holes, string balls,  and sphalerons in proton-proton collisions at \texorpdfstring{$\sqrt{s} = 13\TeV$}{sqrt(s) = 13 TeV}}

\date{\today}

\abstract{
A search for microscopic black holes, string balls, and electroweak sphalerons using proton-proton collisions at $\sqrt{s} = 13\TeV$ recorded with the CMS detector at the CERN LHC during the 2016--2018 data taking, and corresponding to an integrated luminosity of 138\fbinv, is presented. Two search strategies based on control samples in data are used. Model-independent limits on the cross section of physics phenomena with multiple energetic jets, leptons, and photons are set using a method that relies on the shape invariance of the scalar sum of the transverse momenta of all objects in the event. Model-dependent limits on black hole and sphaleron production are set using a newly introduced method that has been developed for the identification of collider events with distinct kinematic features by separating them into classes based on phase space proximity. In the context of models with large extra dimensions, semiclassical black holes and string balls with masses below 8.4--11.4\TeV and 9.0--10.7\TeV, respectively, are excluded at 95\% confidence level, significantly extending the reach beyond previous searches. Results of a dedicated search for electroweak sphalerons are used to derive an upper limit of 0.0034 at 95\% confidence level on the fraction of quark-quark interactions, with a center-of-mass energy above the nominal sphaleron transition energy threshold of 9\TeV, that result in a sphaleron transition.
}

\hypersetup{
pdfauthor={CMS Collaboration},
pdftitle={Search for microscopic black holes, string balls, and sphalerons in proton-proton collisions at sqrt(s) = 13 TeV},
pdfsubject={CMS},
pdfkeywords={CMS, black holes, sphalerons, sphericity, phase space distance, SVM}
}

\maketitle

\section{Introduction}

This paper describes a search for three theoretical objects predicted to exist at the \TeV scale: microscopic black holes (BHs), string balls (SBs), and sphalerons. Microscopic BHs are the manifestation of certain theoretical scenarios beyond the standard model (BSM) that could address open questions of the standard model (SM), such as the hierarchy problem. The Arkani-Hamed--Dimopoulos--Dvali~(ADD)~\cite{Antoniadis:1998ig,Arkani-Hamed:1998jmv,Arkani-Hamed:1998sfv}  and Randall--Sundrum~(RS)~\cite{Randall:1999vf,Randall:1999ee} models address the hierarchy problem by means of extra dimensions with properties that can explain gravity's apparent weakness in the standard 3+1 dimensions of space-time. 
In this paper, we present the results of a search for manifestations of the ADD model.
In this framework, the strength of gravity is greatly increased when its extra dimensional components are taken into account. This shifts the fundamental Planck scale (\md) downward to a level near the electroweak (EW) scale. As a result, microscopic BH formation becomes possible for BHs with mass \mbh $\geq$ \md~\cite{Dimopoulos:2001hw,Argyres:1998qn,Giddings:2001bu}. 
In the semiclassical case where \mbh is well above \md, the microscopic BHs quickly evaporate via Hawking radiation~\cite{Hawking:1975vcx} to a large number of energetic particles, such as gluons, quarks, leptons, and photons. If \mbh is close to \md, quantum effects become significant, altering the decays. For example, quantum BHs decay before they thermalize and produce low-multiplicity final states~\cite{Meade:2007sz, Gingrich:2009hj,Calmet:2008dg}. 

Another model based on string theory predicts that in the region between \md and the threshold for BH formation, highly excited long, jagged SB states~\cite{Dimopoulos:2001qe} can be produced with masses of order $\ms/\gs$, where \ms is the string scale and \gs is the string coupling. The transition from SBs to semiclassical BHs occurs for masses above $\ms/\gs^2$. Similar to semiclassical BHs, SB evaporation yields a large number of SM particles.

Sphalerons are unstable solutions of the electroweak field equations corresponding to saddle-point configurations of the electroweak potential \cite{tHooft:1976rip,Klinkhamer:1984di} that allow baryon and lepton number violation, an essential requirement for the matter-antimatter asymmetry in the universe.  In the SM, there are twelve globally conserved currents $J^\mu = \overline{\psi}_\mathrm{L}\gamma^\mu\psi_\mathrm{L}$, each corresponding to the flavor conservation of one of the 12 fundamental fermions. When the conservation is broken by the electroweak chiral anomaly, the integral of the divergence of these currents over space-time produces what is known as the Chern--Simons number~\cite{Klinkhamer:1984di}, \ncs, which characterizes the topological structure of the EW vacuum. Each sphaleron-induced transition is associated with an \ncs, resulting in a violation of baryon~($B$) and lepton~($L$) numbers by $\Delta B = \Delta L = 3 \ncs$, such that $\Delta(B+L)=6\ncs$ and $\Delta(B-L)=0$. The sphaleron potential is a periodic potential with minima (the vacua) characterized by the integer values of \ncs. The energy required to overcome the potential barriers between minima, known as the sphaleron transition energy (\esph), is estimated to be around 9\TeV~\cite{Tye:2015tva}. An anomaly only occurs when there is sufficient energy to overcome the potential barrier.
The minimal final state from a sphaleron transition, corresponding to $\ncs= \pm 1$, consists of 12 fermions: three charged leptons, three neutrinos, and six quarks (one from each generation and color). Additional quark-antiquark pairs and gauge bosons may also be produced in the process, leading to high-multiplicity final states, potentially observable at the CERN LHC~\cite{Tye:2015tva,Ellis:2016ast,Papaefstathiou:2019djz}.

{\tolerance=500
At the LHC, three types of objects discussed above could produce very similar event signatures involving multiple leptons, photons, and jets that are distinct from SM backgrounds.
\par}

The ATLAS Collaboration published a search for BHs~\cite{ATLAS:2015yln} using 3.6\fbinv of data collected at $\sqrt{s} = 13\TeV$ that was recast by Ellis and Sakurai as a search for sphalerons~\cite{Ellis:2016ast}. The ATLAS Collaboration also searched for quantum BHs in lepton+jet final states using a $\sqrt{s} = 13\TeV$ data set corresponding to an integrated luminosity of 140\fbinv~\cite{ATLAS:2023vat}. An analysis by CMS at the same center-of-mass energy, based on 2016 data, corresponding to an integrated luminosity of 35.9\fbinv, was reported in Ref.~\cite{CMS:2018ozv}. The search reported here is an update to the previous CMS analysis. In addition to using a larger data set, we use a novel phase space metric to set model-dependent limits, while also applying the shape invariance approach used in previous CMS publications~\cite{CMS:2010oej,CMS:2012grf,CMS:2013ndg,CMS:2017boz,CMS:2018ozv}, to set model-independent limits.

The paper is organized as follows. The CMS detector and event reconstruction are presented in Section~\ref{sec:detDesc}.  
Section~\ref{sec:signals} describes the signal models considered and the associated simulated data samples.
Section~\ref{sec:reco} discusses the event reconstruction, while Section~\ref{sec:observables} introduces the variables used in this search.  The event selection is detailed in Section~\ref{sec:presel}, followed in Section~\ref{sec:bkgpred} by descriptions of the two background estimation methods we employ.
Section~\ref{sec:systematics} presents the systematic uncertainties associated with the signal simulation, 
and the results are shown in Section~\ref{sec:results}. The paper is summarized in Section~\ref{sec:summary}.

Tabulated results are provided in the HEPData record for this analysis~\cite{hepdata}.

\section{The CMS detector}
\label{sec:detDesc}

The central feature of the CMS apparatus is a superconducting solenoid of 6\unit{m} internal diameter, providing a magnetic field of 3.8\unit{T}. 
Within the solenoid volume are a silicon pixel and silicon strip tracker, a lead tungstate crystal electromagnetic calorimeter (ECAL), and a brass and scintillator hadron calorimeter (HCAL), each composed of a barrel and two endcap sections. 
Forward calorimeters extend the pseudorapidity ($\eta$) coverage provided by the barrel and endcap detectors. 
Muons are measured in gas-ionization detectors embedded in the steel flux-return yoke outside the solenoid. 
A more detailed description of the CMS detector, together with a definition of the coordinate system used and the relevant kinematic variables, can be found in Refs.~\cite{CMS:2008xjf,CMS:2023gfb}.

The silicon tracker used in 2016 measured charged particles within the range $\abs{\eta} < 2.5$. For nonisolated particles of transverse momenta (\pt) between 1 and 10 \GeV, and $\abs{\eta} < 1.4$, the track resolutions were typically 1.5\% in \pt and 25--90 (45--150)\mum in the transverse (longitudinal) impact parameter~\cite{CMS:2014pgm}. At the start of 2017, a new pixel detector was installed~\cite{Phase1Pixel}; the upgraded tracker measured particles up to $\abs{\eta} = 3.0$ with typical resolutions of 1.5\% in \pt and 20--75\mum in the transverse impact parameter~\cite{DP-2020-049} for nonisolated particles of $1 < \pt < 10\GeV$. According to simulation studies~\cite{DP-2017-015}, similar improvements are expected in the longitudinal direction.

Events of interest are selected by a two-level trigger system. 
The first level, composed of custom hardware processors, uses information from the calorimeters and muon detectors to select events 
at a rate of approximately 100\unit{kHz} within a fixed latency of approximately 4\mus~\cite{CMS:2020cmk}. 
The second level, known as the high-level trigger (HLT), consists of a farm of processors running a version of the full event reconstruction software optimized for fast processing. It reduces the event rate to a few \unit{kHz} before data storage~\cite{CMS:2016ngn,CMS:2024aqx}.

\section{Data and simulated samples}
\label{sec:signals}

The analysis is based on a data sample collected with the CMS detector from 2016 to 2018 in proton-proton ($\Pp\Pp$) collisions at a center-of-mass energy of 13\TeV, corresponding to an integrated luminosity of 138\fbinv.

The BH signal samples were produced at leading order using Monte Carlo (MC) simulation with the \textsc{BlackMax}~v2.02.0~\cite{Dai:2007ki} 
and \textsc{Charybdis2}~v1.003~\cite{Harris:2003db,Frost:2009cf} generators at $\sqrt{s} = 13\TeV$. The \textsc{Charybdis2} generator models BH decay primarily through semiclassical Hawking radiation with graybody factors, while \textsc{BlackMax} offers more flexible parameterizations for the decay spectra and different treatments of mass and angular momentum loss during evaporation.
The SB samples were generated using \textsc{Charybdis2}.
The sphaleron samples used the \textsc{BaryoGEN}~v1.0~\cite{Bravo:2018dkg} generator. Simulations for quantum chromodynamics (QCD) multijet production with \MGvATNLO~v2.2.2~\cite{Frederix:2018nkq} at leading order using the MLM matching scheme~\cite{Alwall:2007fs} have been employed to guide the selection criteria and train machine-learning models, as detailed in Section~\ref{sec:presel}.

The BH samples were produced under different assumptions regarding the angular momentum, energy loss, and decay mechanisms.
Samples labeled with ``B'' were produced using the \textsc{BlackMax} generator:

\begin{itemize}
    \item B1: nonrotating;
    \item B2: rotating with no graviton emission included;
    \item B3: rotating with energy and angular momentum loss (graviton emission included);
\end{itemize}

while those labeled with ``C'' were produced using the \textsc{Charybdis2} generator:
\begin{itemize}
    \item C1: rotating;
    \item C2: nonrotating;
    \item C3: rotating with the evaporation model in Ref.~\cite{Creek:2007sy,Creek:2007pw};
    \item C4: rotating with the Yoshino--Rychkov (YR) loss model~\cite{Yoshino:2005hi}, which is a conservative bound that includes impact parameter considerations; 
    \item C5: rotating, stable-remnant model~\cite{Koch:2005ks,Stoecker:2006we,Scardigli:2008jn}, where the stable remnant must be a colorless object essentially equivalent to a quark-antiquark bound state;
    \item C6: rotating, boiling remnant model~\cite{Dimopoulos:2001qe,Gingrich:2008di}, where the remnant continues to evaporate until a maximum Hawking temperature equal to \md is reached.
\end{itemize}

Furthermore, the samples were simulated with various combinations of model parameters. For each model, the fundamental Planck scale \md is varied between 2 and 9\TeV in 1\TeV steps and for every mass value, n = 2, 4, and 6 extra dimensions are used. 
The number of extra dimensions corresponds to representative benchmarks commonly used in previous searches and spans a broad range of phenomenologically relevant models. 
The parameter grids are chosen to provide adequate coverage of the accessible phase space while maintaining a manageable number of simulated samples. 
The signal model varies smoothly between neighboring simulated mass points, making interpolation between them appropriate. 
This procedure was validated by comparing limits obtained from interpolated signal templates at intermediate mass points with those obtained by interpolating between limits at neighboring simulated mass points, yielding very similar results.
As BH production is a threshold phenomenon, we search for BHs above a certain minimum mass \mbhmin. We vary \mbhmin in simulation from $(\md + 1)\TeV$ to 12\TeV in 1\TeV steps.

The production of SBs depends on two parameters, the string scale \ms and the string coupling \gs. For all SB samples, $n=6$ is used.
In weakly-coupled string scenarios compactified to $4+n$ dimensions, \md and the string parameters are related by \cite{Gingrich:2008di},
\begin{equation}
    \md = \gs^{-(\frac{2}{n-2})} \ms.
    \label{eq:gsCalc}
\end{equation}
In simulation, the string scale \ms is varied from 1 to 4\TeV in 0.5\TeV steps. For each mass value, \gs values of 0.2, 0.3, and 0.4 are used, while the minimum mass of SBs ranges from 7 to 12\TeV in steps of 1\TeV. These parameter ranges cover both the formation of SBs and their transition to semiclassical BHs.

Sphaleron production depends on two parameters, \esph and \ncs. The parameter \esph is scanned in a range from 8.5 to 9.5\TeV in steps of 0.25\TeV. 
Since the sphaleron transition is exponentially suppressed by $\abs{\ncs}$, only $\ncs= \pm 1$ are simulated. The probability of $\ncs = +1$ is denoted as \pncs. The \pncs values of 0\%, 50\%, and 100\% were used, where, for example, $\pncs =50\%$ means half of the events have $\ncs = +1$ and half have $\ncs = -1$. 

The theoretical cross section for sphaleron production is calculated in terms of the partonic cross section~\cite{Papaefstathiou:2019djz},
\begin{equation}
    \hat{\sigma}(\sqrt{\hat{s}}) = \frac{\psph}{m_\PW^2} \Theta(\sqrt{\hat{s}} - \esph),
    \label{eq:sphaleron_xs}
\end{equation}
where $m_\PW$ is the mass of the \PW\ boson, $\hat{s}$ is the square of the partonic center-of-mass energy, and $\Theta$ is the Heaviside step function. The partonic cross section is parameterized by the two parameters \esph and \psph, where the latter is a dimensionless phenomenological pre-exponential factor that scales the reference sphaleron production cross section above the threshold \esph. 
The sphaleron samples were generated assuming $\psph = 1$, which facilitates setting a limit on this factor.

In all MC samples, \PYTHIA~v8.205 was used for hadronization with tune CP5~\cite{CMS:2019csb}. The parton distribution function (PDF) set used in simulation is NNPDF3.1 at next-to-next-to-leading order in perturbative QCD~\cite{NNPDF:2017mvq}. 
Simulated minimum bias events are superimposed on the primary collision to mimic the impact of additional interactions that occur in the same or adjacent bunch crossings (pileup)~\cite{Lange:2159066}.

\section{Event reconstruction}
 \label{sec:reco}

For this analysis, which focuses on searching for signals at the \TeV scale, events are selected using an HLT criterion based on the scalar sum of transverse momenta of the reconstructed jets  (\HT) in the event. The lowest \HT threshold that ensures optimal performance and data coverage is chosen. For data taken in the year 2016 this threshold is 900\GeV, while for 2017--2018 it is 1050\GeV. The resulting trigger selection is fully efficient for events that subsequently satisfy the offline requirements used in the analysis. Beyond the HLT requirement, a preselection of $\st > 2\TeV$ is applied, where \st is defined as the scalar sum of the transverse momenta of all reconstructed objects in the event. The subsequent requirements described in this section define the full event selection used in the analysis, for which particles are reconstructed using the particle-flow (PF) algorithm~\cite{CMS:2017yfk}. 

The PF algorithm aims to identify each individual particle in an event, with an optimized combination of information from the various elements of the CMS detector.  The energy of photons is obtained from the ECAL measurement. 
The energy of electrons is determined from a combination of the electron momentum at the primary interaction vertex as determined by the tracker, 
the energy of the corresponding ECAL cluster, and the energy sum of all bremsstrahlung photons spatially compatible with emission from the electron along its curved track.
The energy of muons is obtained from the curvature of the corresponding track. 
The energy of charged hadrons is determined from a combination of their momentum measured in the tracker and the matching ECAL and HCAL energy deposits, 
corrected for the response function of the calorimeters to hadronic showers. 
Finally, the energy of neutral hadrons is obtained from the corresponding corrected ECAL and HCAL energies.

The primary vertex is taken to be the vertex corresponding to the hardest scattering in the event, evaluated using tracking information alone, as described in Section 9.4.1 of Ref.~\cite{CMS-TDR-15-02}.
Events are required to contain a primary vertex within 24\unit{cm} of the nominal collision point along the beam axis and within 2\unit{cm} in the transverse direction.

Jets are clustered from PF particles using the anti-\kt algorithm with a distance parameter $R = 0.4$. 
The jet momentum is determined as the vectorial sum of all particle momenta in the jet, and is found from simulation to be within 5 to 10\% of the true momentum over the whole \pt spectrum and detector acceptance.
Jet energy corrections are derived from simulation to bring the average measured jet energy into agreement with that of particle-level jets.
In situ measurements of the momentum balance in dijet, $\PGg$+jets, and $\PZ$+jets events are used to determine any residual differences between the jet energy scale in data and in simulation, and appropriate corrections are applied~\cite{CMS:2016lmd}.
Pileup interactions can contribute additional tracks and calorimetric energy depositions to the jet momentum.
To mitigate this effect, charged particles matched to pileup vertices are removed and an offset correction is applied to correct for remaining neutral-particle contributions~\cite{CMS-PAS-JME-14-001}.
All jets are required to have $\pt > 70\GeV$, and pseudorapidity $\abs{\eta} < 2.5$. The threshold of $70\GeV$ is chosen consistently for all reconstructed objects to ensure sufficient signal acceptance while preserving sensitivity across a wide range of signal samples.
Additional selection criteria~\cite{CMS:2017wyc} are applied to each jet to remove those originating from calorimetric noise.
An event-level filter is applied such that the energy fraction carried by muon candidates in the leading \pt jet that fails the standard identification must be less than 80\%. This requirement removes events in which a low-momentum muon is misreconstructed with very high momentum and misidentified as a high-energy jet.

Electrons are reconstructed as charged-particle tracks with additional energy depositions in the ECAL~\cite{CMS:2015xaf}. 
A requirement of pseudorapidity $\abs{\eta} < 2.5$ is applied to all electrons. 
Identification criteria make use of a discriminant that combines information on the reconstructed electron and its isolation from other activity in the event. Additional constraints on the transverse and longitudinal impact parameters of electron tracks ensure consistency with the primary vertex. 
The chosen discriminant working point corresponds to a selection efficiency of approximately 80\%~\cite{CMS:2020uim}. 

Photon reconstruction is based on information from the electromagnetic calorimeter with the requirement of no matched charged-particle tracks. 
As for electrons, a requirement of 
pseudorapidity $\abs{\eta} < 2.5$ is applied along with an identification discriminant working point that achieves approximately 80\% efficiency~\cite{CMS:2020uim}.

{\tolerance=500
Muon candidates are reconstructed using the PF algorithm and are categorized as global muons~\cite{CMS:2012nsv}, by the combined fit of tracks in the silicon tracker and the muon system. Tight selection criteria are then applied to remove misidentified muons from hadronic decays or hadron shower remnants~\cite{CMS:2018rym}.
An isolation requirement is applied to reduce contamination from nonprompt muons originating from decays of hadrons within jets.
Muon candidates are required to have pseudorapidity $\abs{\eta} < 2.4$.
\par}

An object arbitration procedure is used to disambiguate objects in the event. It is applied to photons and jets using a cone centered on their reconstructed approximate trajectories, of size $\Delta R = \sqrt{\smash[b]{(\Delta\eta)^2+(\Delta\phi)^2}} = 0.4$.
The arbitration proceeds in a specific hierarchical order as follows.
First, all leptons are retained. 
Next, photons are selected only if there is no lepton in the cone. 
Finally, jets are selected from the remaining objects, provided that there is no lepton or photon in the cone.

The missing transverse momentum vector \ptvecmiss is defined as the projection onto the plane perpendicular to the beam axis of the negative vector sum of the momenta of all reconstructed PF objects in an event~\cite{CMS:2019ctu}. Its magnitude is referred to as \ptmiss. The jet energy corrections are propagated into the \ptmiss reconstruction.

\section{Observables}
\label{sec:observables}

The key observables used in this analysis to discriminate between signal and background include the scalar sum of transverse momenta of all objects in the event (\st), sphericity ($S$)~\cite{ALEPH:1996oqp,ATLAS:2012tch,ALICE:2012cor}, and in the case of the model-dependent search, a novel phase space distance metric is also employed. The objects in an event are jets, electrons, photons, muons, and missing transverse momentum (\ptmiss).

\subsection{Scalar sum of transverse momenta}
The \st variable is defined as the scalar sum of the \pt of all objects in the event,
\begin{equation}
    \st = \left( \sum_{i=1}^{N} {\pt}_i \right) + \ptmiss .
    \label{eq:stdef}
\end{equation}
The \st distributions for various signal models, together with that of the QCD multijet background, are shown in Fig.~\ref{fig:st_and_sphericity} (left). Among BH signals, we use the B1 model with $n=2$ and $\md =$~2$\TeV$ as a representative case for the description of the analysis, including many figures and resulting exclusions. Where relevant, figures focus on the \mbh and \md mass points near exclusion. Other models are treated similarly and the limits associated with them are included in the final exclusion plots in section \ref{sec:results}.  For the sphaleron signal, the nominal $\esph = 9\TeV$ with different \pncs assumptions is shown.

The shape of the \st distribution for background processes, which is dominated by QCD multijet background, is assumed to be independent of the object multiplicity, $N$. 
This facilitates the modeling of the background from control regions (CRs) and the extension of its prediction to higher multiplicities where BSM physics signals may be present. This is discussed in more detail in Section~\ref{subsec:SI_method}.

\subsection{Sphericity}
\begin{figure}
    \centering
    \includegraphics[width=0.49\textwidth]{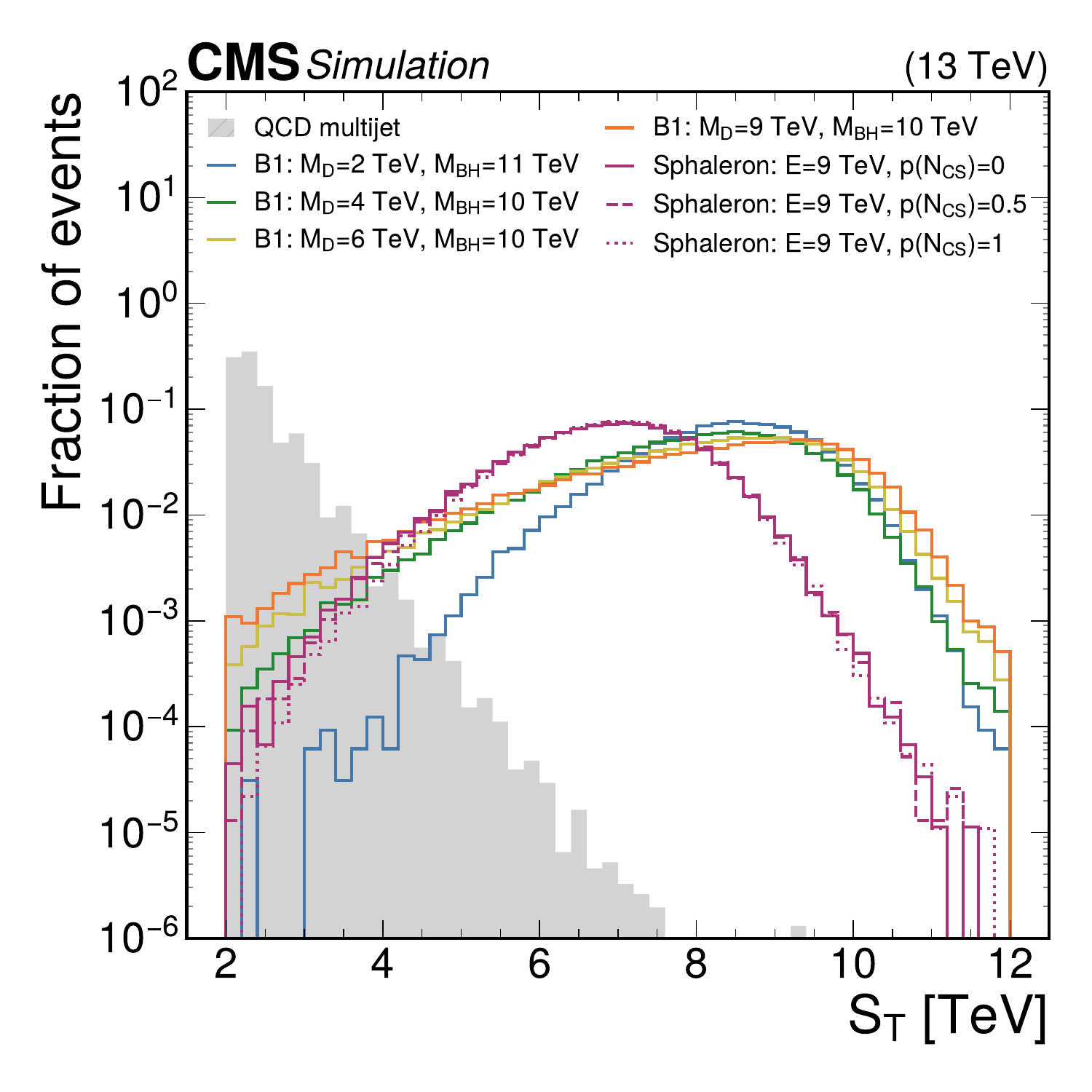}
    \includegraphics[width=0.49\textwidth]{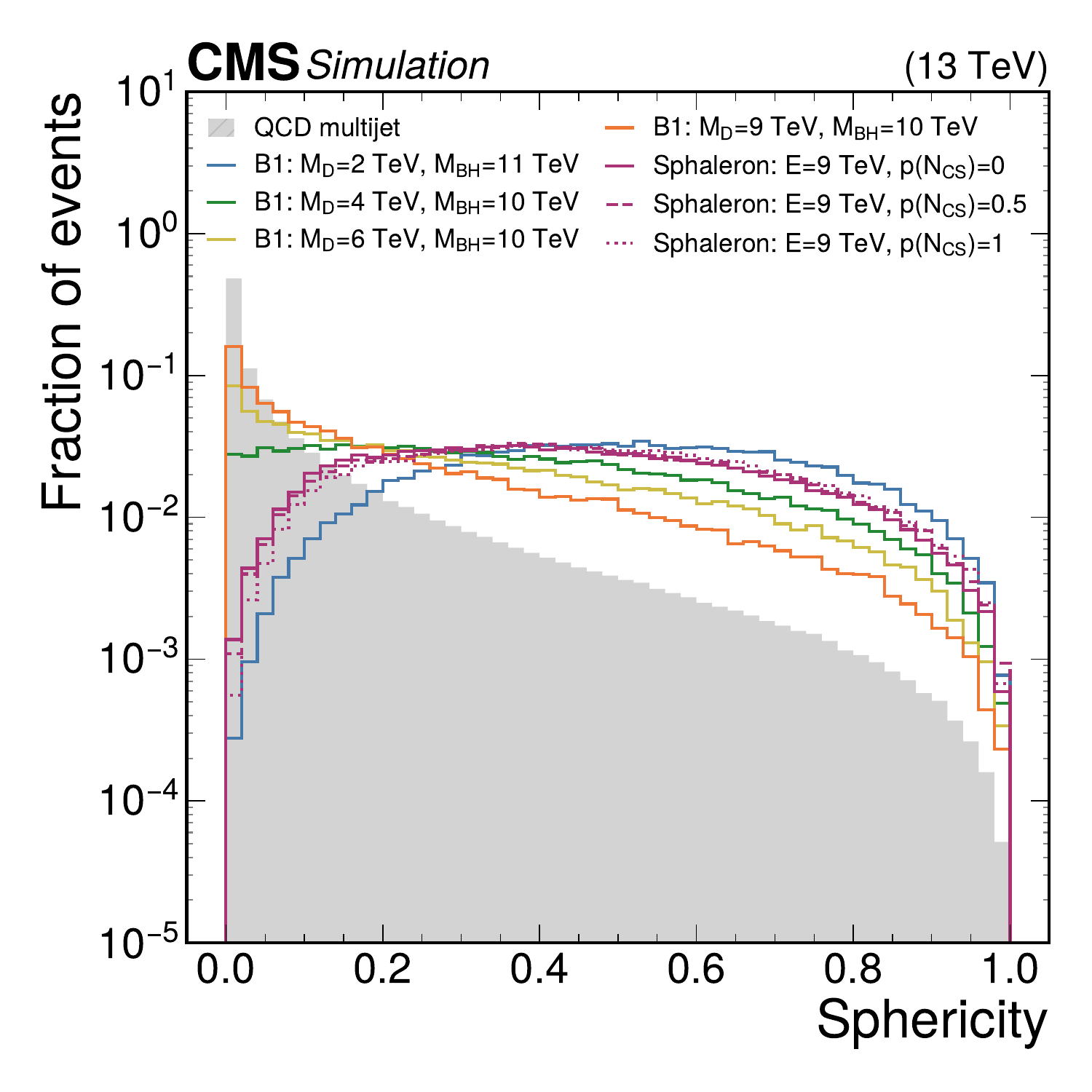}
    \caption{The \st (left) and sphericity (right) distributions for various BH (with $n = 2$) and sphaleron signal models are plotted along with the corresponding distributions for  simulated QCD multijet background events. The distributions are normalized to unit area.}
    \label{fig:st_and_sphericity}
\end{figure}

Sphericity is a measure of how isotropic an event is, and it is employed to improve signal sensitivity.
In this paper, sphericity is defined as follows:
\begin{equation}
    S = \frac{2 \lambda_2}{\lambda_2 + \lambda_1},
    \label{eq:sphericity}
\end{equation}
where $\lambda_1$ and $\lambda_2$ are the eigenvalues of the transverse momentum tensor,
\begin{equation}
    S_{xy}^L = \frac{1}{\sum_i {\pt}_i} \sum_i \frac{1}{{\pt}_i} 
    \begin{bmatrix}
    p_{xi}^2 & p_{xi} p_{yi} \\
    p_{xi} p_{yi} & p_{yi}^2
    \end{bmatrix},
    \label{eq:Sxymatric}
\end{equation}
where the index $i$ extends over all objects passing our selection criteria. The definition of sphericity in this paper is restricted to the transverse plane and has been referred to as transverse sphericity~\cite{ATLAS:2012tch,ALICE:2012cor}. 
Signal events tend to be more spherical than background events due to the isotropic decays of BHs and sphalerons. Therefore, it is expected that the sphericity distribution for signal events will be shifted towards higher values compared to the QCD multijet background, as shown in Fig.~\ref{fig:st_and_sphericity} (right). 
As the mass of the BH increases, its temperature decreases, which results in a larger number of softer objects emitted in the process of evaporation. BH models with higher \mbh tend to have a more pronounced shift to higher sphericity values.

\subsection{Phase space distance}
Novel representations of collider data play an increasingly important role in modern machine learning methods. 
For example, optimal transport distances using energy flow distributions, have given rise to a new geometry of collider events~\cite{OT_paper,OT_geometry,OT_linear}
that has widespread applications in particle physics~\cite{Komiske:2019jim,Cesarotti:2020hwb,Cesarotti:2020ngq,Gouskos:2022gjg,ATLAS:2023mny}.
However, less attention has been given to the intrinsic geometry of the phase space manifold where scattering amplitudes are defined.
A natural distance between collider events with $N$ objects was proposed in Ref.~\cite{Larkoski:2020thc}. To this end, a covariant description of the massless $N$-body phase space manifold $\Pi_N$ is presented and shown to be isomorphic to the product space of an ($N-1$)-simplex $\Delta_{N-1}$ and a ($2N-3$)-hypersphere $S_{2N-3}$: $\Pi_N \cong \Delta_{N-1}\otimes S_{2N-3}$. Explicit and global coordinates of the phase space, ${\vec \rho}$ for the simplex and ${\vec v}$ for the hypersphere, are obtained via   transforms of the four-vectors of the $N$ objects in the event.

A practical metric for collider events using global phase space coordinates was developed and tailored to the structure of events at hadron colliders in the center-of-mass frame~\cite{Cai:2024xnt}.
After pre-processing the collider events and the manifold, the metric distance between two events, $\mathrm{A}$ and $\mathrm{B}$, is defined by
\begin{equation}
    d_{\Pi}({\vec \rho}_\mathrm{A},{\vec v}_\mathrm{A}'; {\vec \rho}_\mathrm{B},{\vec v}_\mathrm{B}') = 
    \sqrt{ \frac{1}{16\pi^2}\left( \frac{c}{4} \right)^{\frac{3-2N}{3N-4}} d^2_\Delta({\vec \rho}_\mathrm{A}, {\vec \rho}_\mathrm{B})
           + \frac{1}{4\pi^2}\left( \frac{c}{4} \right)^{\frac{N-1}{3N-4}} d^2_\mathrm{S}({\vec v}_\mathrm{A}', {\vec v}_\mathrm{B}') },
    \label{eq:distance_total}
\end{equation}
where $d_\Delta$ is the distance on the simplex, $d_\mathrm{S}$ is the distance on the hypersphere, and $c$ serves as a relative reweighting factor that gives rise to equally valid phase space distances. The relative reweighting factor is optimized to maximize the performance of the phase space distance in distinguishing signal from background events. In this analysis, we use the ratio sphere\,:\,simplex $=4:1$, defined as the ratio of the coefficients that multiply the hypersphere term $d^2_\mathrm{S}$ and the simplex term $d^2_\Delta$ in Eq.~(\ref{eq:distance_total}). For any object multiplicity $N$ this ratio of coefficients reduces to $c$, so the choice $4:1$ corresponds to setting $c=4$.
The definitions of the coordinates and derivations of the distances can be found in Ref.~\cite{Cai:2024xnt}.
When combined with downstream machine-learning algorithms, the phase space distance achieves approximately 80\% accuracy in distinguishing top quark pair production events from QCD dijet events in MC simulations.

Given that the dimension of the phase space manifold depends on the particle multiplicity $N$, a consistent means of defining the distance between events with different multiplicities is needed. In this work, we achieve this by fixing the number of objects per event to $N=30$. This number covers the highest multiplicity seen in signal MC samples. When fewer than 30~objects are available in an event, the remaining entries are `zero-padded', \ie, they are assigned a value of zero for subsequent calculations.

\section{Event selection}
 \label{sec:presel}

A sphericity requirement of $S > 0.1$, is applied to reject background events, particularly at low \st, where background contamination is highest.
The sphericity requirement is determined from a scan of the sphericity threshold, evaluating the expected significance for all signal models considered. With the $S > 0.1$ selection, an average improvement in sensitivity of approximately 70\% is observed across the signal models used in this search. While a small number of models show a mild reduction in significance, the majority of signal models have a substantial gain, with improvements up to $\sim$200\%.

A preselection requirement of $N \geq 3$ objects is applied for the background fit in the model-independent search described in Section~\ref{subsec:SI_method}, while $N \geq 4$ objects is required for the phase space distance method described in Section~\ref{subsec:PS_method}.

Due to the high dimensionality of the phase space manifold, the phase space distance, defined in Section~\ref{sec:observables}, is not an optimal discriminator on its own and so it is combined with a machine-learning classifier to enhance the separation between signal and background.
The support vector machine (SVM)~\cite{Cortes:1995hrp} classifier is chosen for this purpose because of its efficacy in handling high-dimensional data and because of its intrinsic design that makes it a natural choice for processing distance-based inputs.
Using pairwise phase space distances as input, the SVM classifies events by constructing a maximum-margin hyperplane that maximizes the distance to the nearest data points from each class.
This approach is more interpretable than a ``black-box" neural network because the decision boundary (hyperplane) is constructed with physically motivated phase space distances. The radial basis function is used as the kernel to provide a non-linear generalization for the SVM classifier,
\begin{equation}
    k(\vec{x}_\mathrm{A};\vec{x}_\mathrm{B})=\exp\left[-\gamma d(\vec{x}_\mathrm{A};\vec{x}_\mathrm{B})^2\right],
    \label{eq:rbf_kernel}
\end{equation}
where $d(\vec{x}_\mathrm{A};\vec{x}_\mathrm{B}) = d_{\Pi}({\vec \rho}_\mathrm{A},{\vec v}_\mathrm{A}'; {\vec \rho}_\mathrm{B},{\vec v}_\mathrm{B}')$ is the phase space distance between event $\mathrm{A}$ and event $\mathrm{B}$, and $\gamma$ is a hyperparameter that controls the width of the kernel, which is optimized during the validation step of the training.
The SVM model provides a probability for how signal-like an event is, which is used as the SVM score in the subsequent analysis. To train the SVM, a simulated QCD multijet sample is used for background, and a mixture of mass points in the nonrotating B1 model with $n = 2$ extra dimensions is used as signal. This general model was evaluated on other signals, including sphalerons, and demonstrated good performance. This is likely because the B1 model with $n = 2$ is the most background-like, and so cases where the signal is further from the background in this metric are less challenging to distinguish.

\begin{figure}
    \centering
    \includegraphics[width=0.49\textwidth]{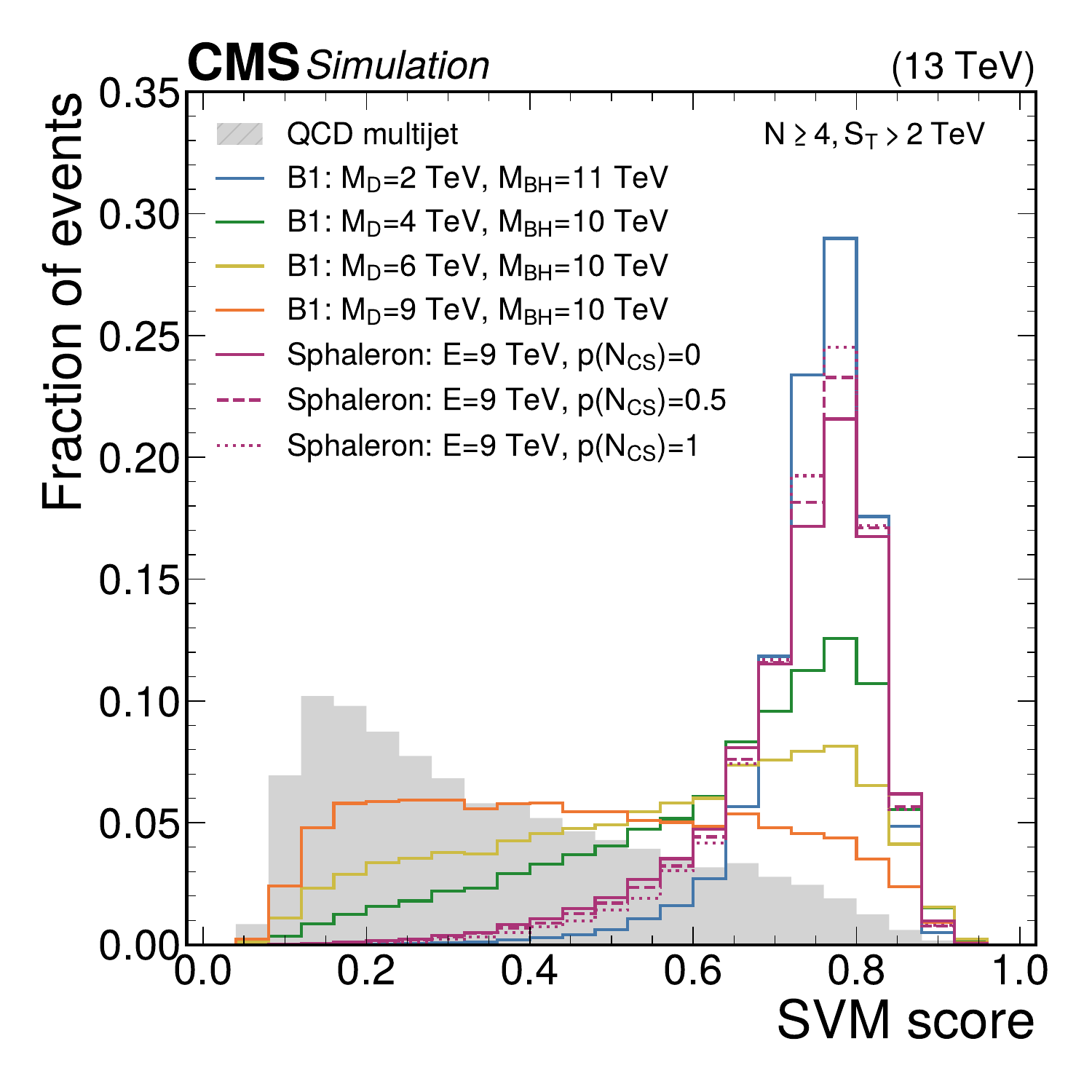}
    \includegraphics[width=0.49\textwidth]{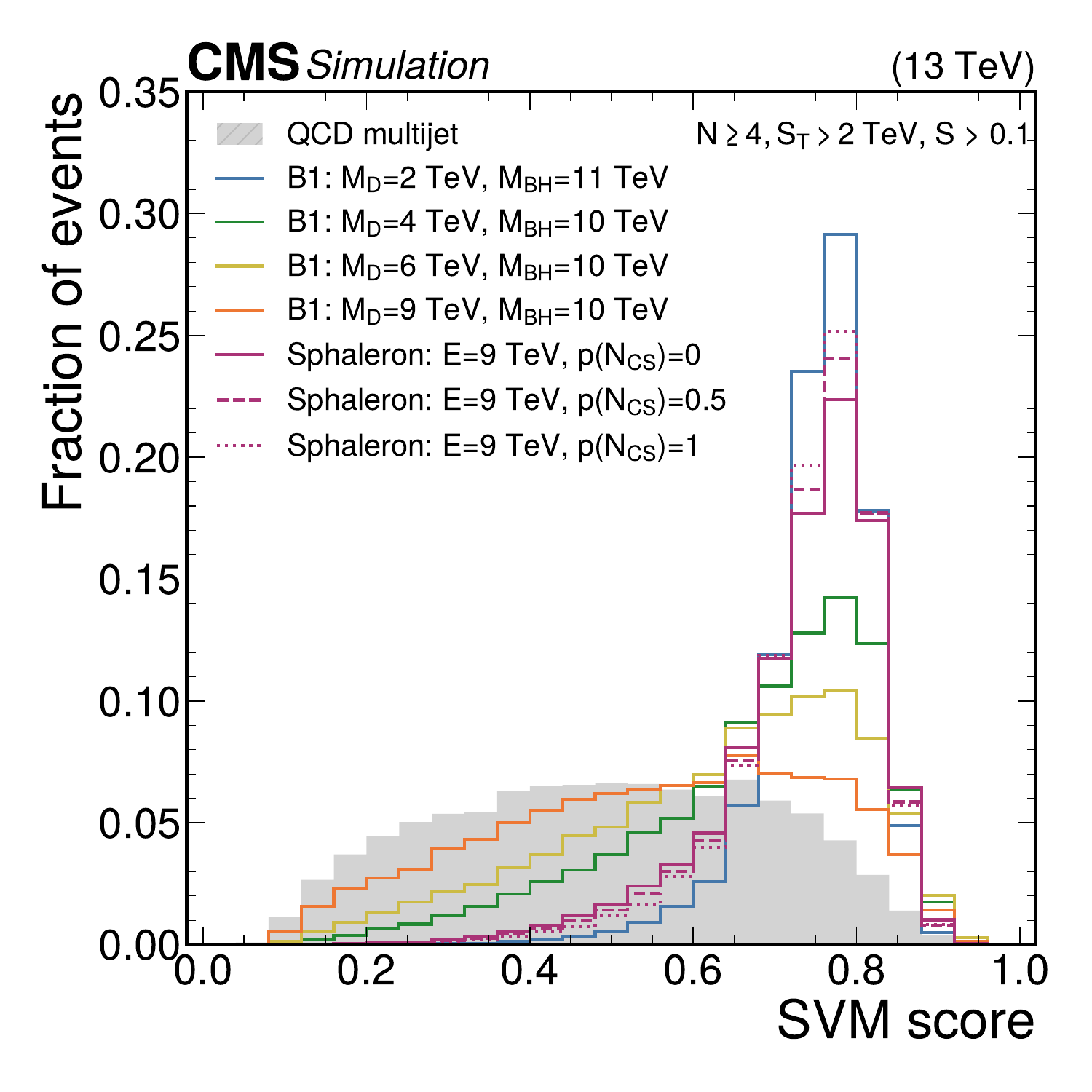}
    \caption{The SVM score distributions for simulated QCD multijets, and selected BH (with $n = 2$) and sphaleron models, before (left) and after (right) the sphericity requirement. The distributions are normalized to unit area.}
    \label{fig:svm_with_wo_sphericity}
\end{figure}

Figure~\ref{fig:svm_with_wo_sphericity} shows the SVM score distributions normalized to equal area for simulated QCD multijet background and selected signal models, before (left) and after (right) the sphericity requirement.

The value of the SVM score used to define the final signal search region (SR) in data is optimized to maximize signal sensitivity as measured by the Punzi figure of merit~\cite{Punzi:2003bu} with $\sigma = 5$ and is found to be 0.63 for several signal models. To evaluate the stability of the results near this choice, the threshold was shifted to 0.5 or 0.7 resulting in a 4~to~25\% variation in the yield of signal events, depending on the mass point.

\begin{figure}
    \centering
    \includegraphics[width=0.49\textwidth]{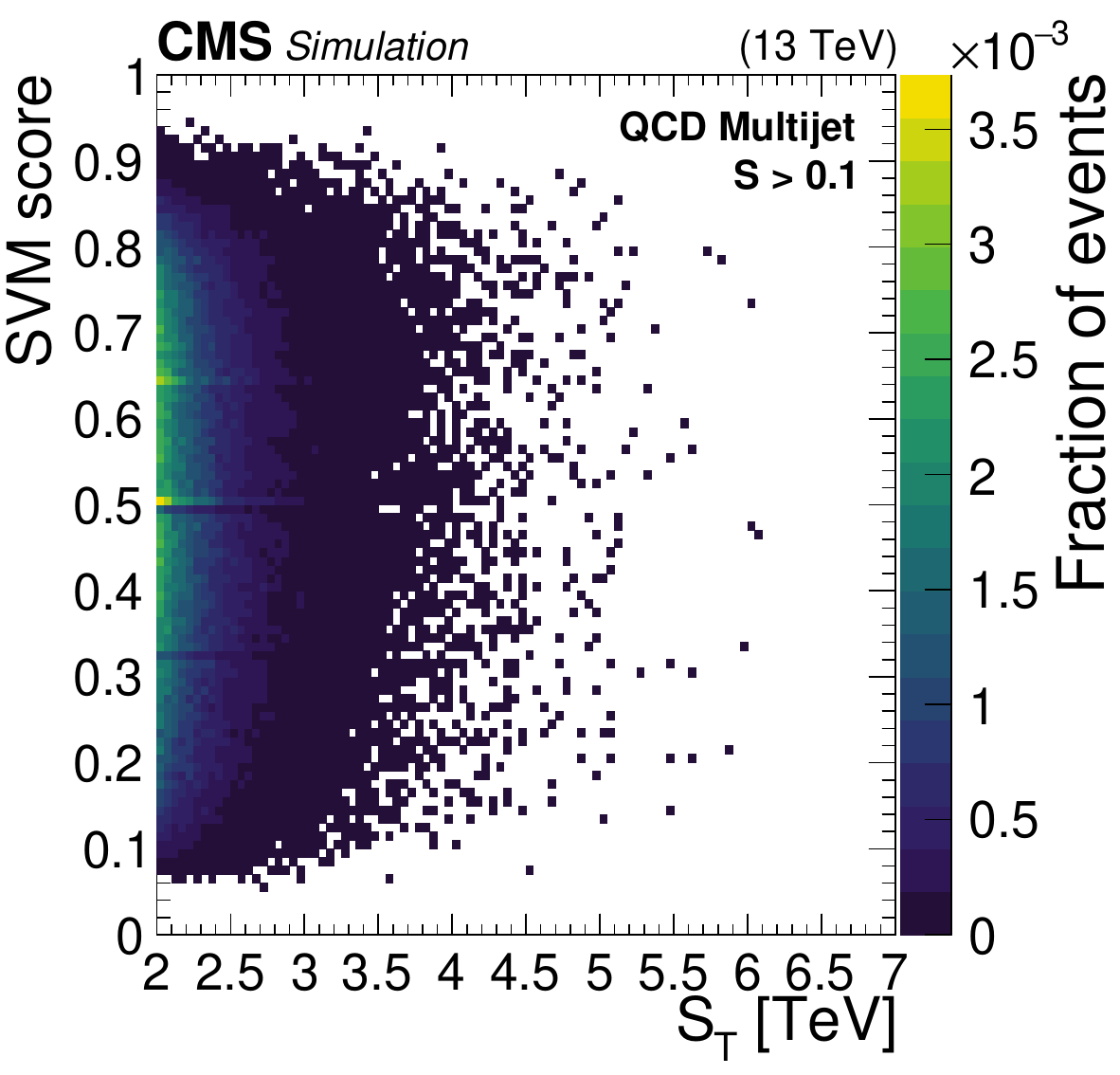}
    \includegraphics[width=0.49\textwidth]{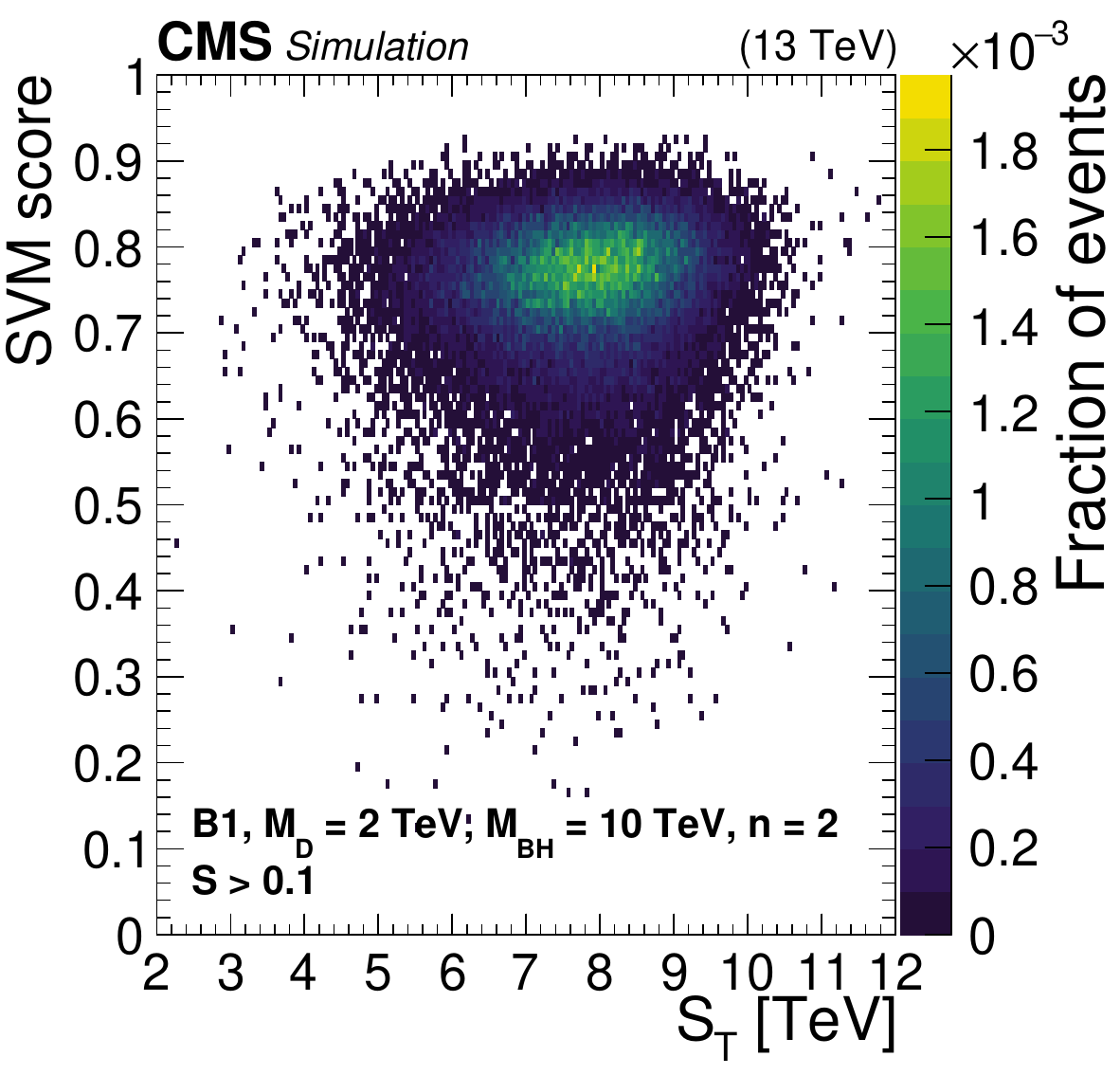}
    \caption{The SVM score vs. the \st distributions for simulated QCD multijet background (left) and the BH signal model B1 with $\md = 2\TeV$, $\mbh = 10\TeV$, and $n = 2$ (right), after the $S>0.1$ selection.}
    \label{fig:svm_vs_st}
\end{figure}

Figure~\ref{fig:svm_vs_st} shows distributions of  SVM score versus \st  for simulated QCD multijet background events (left) and the BH signal model B1 with $\md = 2\TeV$, $\mbh = 10\TeV$, and $n = 2$ (right). Clear separation between signal and background is observed.

\section{Background prediction methods}
\label{sec:bkgpred}

The dominant background is expected to originate from QCD multijet events. The contribution from processes, like \ttbar, $\PW$+jets or $\PZ$+jets, is found to be negligible. As noted above, the searches presented in this paper follow two different approaches. The two approaches also entail different background prediction methods.

\subsection{Shape invariance method}
\label{subsec:SI_method}

The shape invariance (SI) approach, following previous CMS publications on this topic~\cite{CMS:2010oej,CMS:2012grf,CMS:2013ndg,CMS:2017boz,CMS:2018ozv}, assumes \st invariance with respect to the multiplicity $N$, which allows for an estimate of the QCD background across samples selected with different minimum multiplicities. 
Simulation studies confirm that the shape of the \st distribution remains consistent across multiplicities, allowing the use of low-multiplicity events to model the \st distribution of high-multiplicity backgrounds. 
Deviations associated with initial-state radiation and the selection on sphericity are taken into account by additional systematic uncertainties, as described in Section~\ref{sec:systematics}.

The method proceeds by first fitting 16~analytic functions to the \st distribution from events with multiplicity $N = 3$ in a control region (SI-CR).
These functional forms have been commonly used in previous searches to model the QCD \st spectrum and are known to provide a good description of the data within the fit region.
The SI-CR, which spans from $\st = 2.5\TeV$ to $\st = 5.2\TeV$, is optimized to ensure robust background estimation, while maintaining negligible signal contamination.
To ensure a physically meaningful behavior of the background prediction, functions that do not exhibit a monotonically decreasing behavior over the relevant \st range are excluded. All remaining candidate functions provide an acceptable description of the data in the SI-CR, as determined by goodness-of-fit tests.
The choice of the nominal function is based on a comparison of the integrals of the candidate functions in the range $5 < S_T < 7\TeV$. The function whose integral is closest to the central value, defined as the average of the mean and median of the set of accepted functions, is selected as the nominal function. This function defines the nominal background shape prediction. The spread among the accepted functions is treated as a systematic uncertainty, as described in Section~\ref{sec:systematics}. This function is then propagated to higher multiplicities of the final-state objects. 
The nominal function, chosen to be representative of the central behavior of the accepted function set, is given by
\begin{equation}
    f(\st)
    = \frac{p_0\,\bigl(1 - \bigl(\st'\bigr)^{1/3}\bigr)^{p_1}}
       {\bigl(\st'\bigr)^{\,p_2 + p_3 \ln\!\bigl(\st'\bigr)}},
    \label{eq:bestFit}
\end{equation}
where $\st' = \st / \sqrt{s}$ and $p_i$ ($i = 0,1,2,3$) are the fit parameters. 

\begin{figure}[!htbp]
    \includegraphics[width=0.49\textwidth]{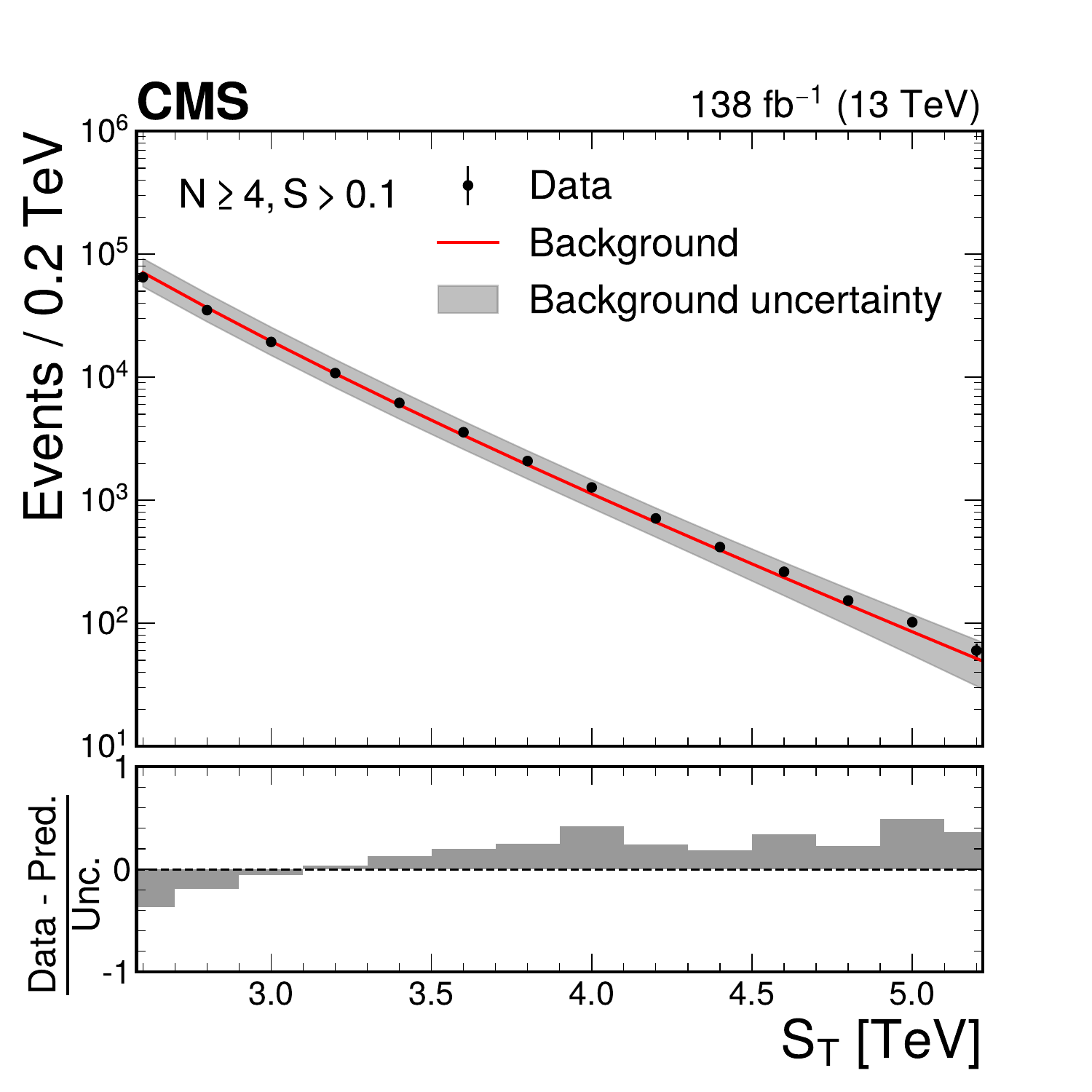}
    \includegraphics[width=0.49\textwidth]{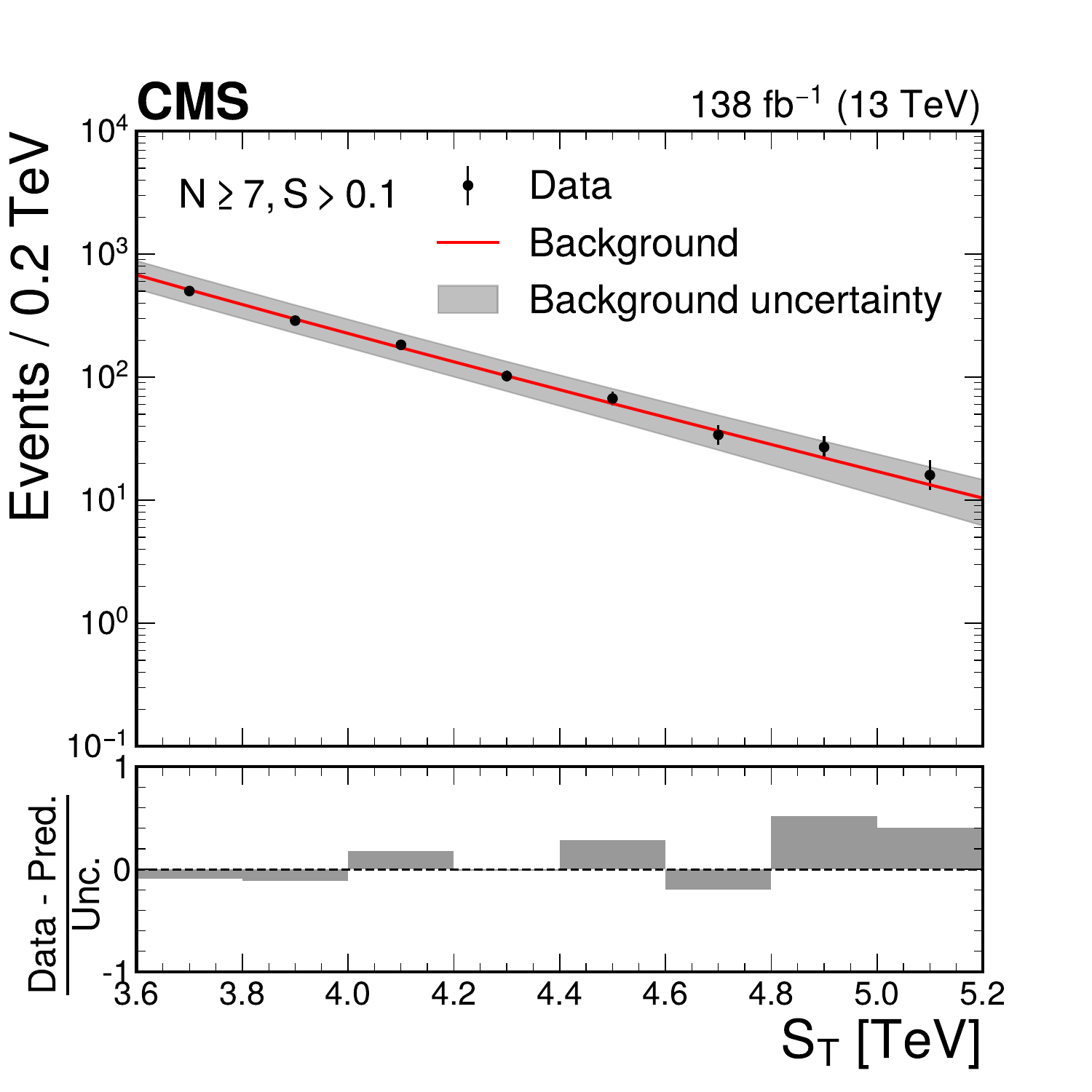}
    \caption{The \st distribution in the $N \geq 4$ (left) and $N \geq 7$ (right) in the SI-VR in data is indicated by the black dots. The background prediction is represented by the red line, and the gray band corresponds to the background modeling uncertainty. The lower panels show the difference between observed data and the background prediction, normalized by the total uncertainty.}
    \label{fig:postfit_SI_VR}
\end{figure}

For each multiplicity threshold, the background estimate 
is normalized within a specific low-\st window. The boundaries of this window are determined using simulated QCD multijet events.
The window begins at the \st value where the ratio relative to the exclusive $N=3$ distribution reaches 99\% of its plateau value. The exclusive $N = 3$ distribution refers to the fitted background function obtained from the 2.5--5.2\TeV  range. The window then extends 400\GeV beyond this threshold, ensuring that there is a sufficient number of events in the normalization region.
The regions of \st between 2.5 and 5.2\TeV then serve as a validation region (SI-VR) to ensure that the background estimation works well in data. The \st range directly above this region defines the signal region (SI-SR).
Figure~\ref{fig:postfit_SI_VR} shows that the prediction in SI-VR describes the data within the uncertainties for different multiplicity thresholds, including both low and high multiplicities.

\subsection{Phase space distance method}
\label{subsec:PS_method}

The background prediction method for the phase space (PS) distance approach is the Alphabet method~\cite{CorcodilosPhD}, which relies on control samples in data as described below.

First, the SVM score is used to define regions enriched in background and depleted of signal. The SR, referred to as the PASS region, contains events with SVM score greater than or equal to 0.63. The CR, also referred to as the FAIL region, is defined by the complementary set of events with a score smaller than 0.63.
All events are required to satisfy $N \geq 4$ and $S > 0.1$ as discussed in the previous section.

Binned \st distributions are then created for both PASS and FAIL regions.
The background yields in the PASS region are estimated from the observed data distribution in the FAIL region, multiplied by a transfer function \RPFj,
 \begin{equation}
	 \NPASS = R_{\text{P/F}}(j) \, \NFAIL ,
	 \label{eq:rpf}
\end{equation}
where \NPASS\ is the estimated background yield in the $j$-th bin of the  \st distribution in the PASS region,
\NFAIL\ is a floating parameter that determines the background yield in the $j$-th bin of the \st distribution in the FAIL region, which is constrained to values close to the observed data yield in the fit described below, and  \RPFj\ is the value of the pass-to-fail ratio for bin $j$. 
The \RPF\ ratio is modeled by an exponential function to ensure a positive yield. Higher-order polynomials were found to have comparable performance when evaluated by the Fisher $F$-test~\cite{WileyChi2,BAKER1984437}.

Finally, a binned maximum likelihood fit is performed simultaneously in the PASS and FAIL regions to extract both the background model, including the parameters of the \RPFj function and the binned background yield in the FAIL region, and the signal strength. 
All correlations between the parameters of the \RPFj function and the nuisance parameters are accounted for by construction.

\begin{figure}[ht!]
    \centering
    \includegraphics[width=0.9\textwidth]{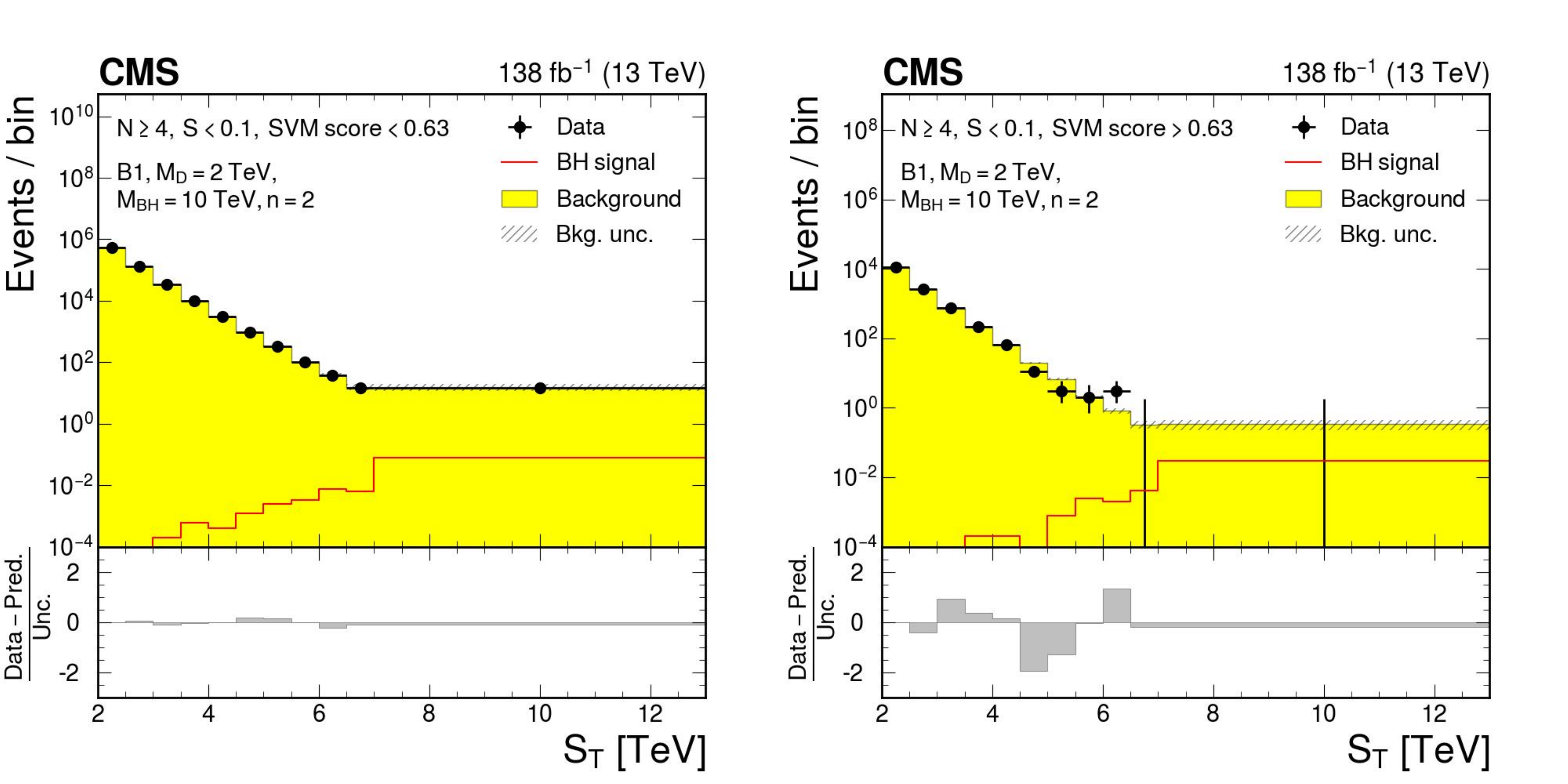}
    \caption{Post-fit \st distributions in VR-FAIL (left) and VR-PASS (right) regions in data. The gray hatched areas include both statistical and systematic uncertainties in the background prediction (yellow histogram). The red line corresponds to the signal model B1, with $\md = 2\TeV$, $\mbh = 10\TeV$, and $n = 2$. The lower panels show the difference between observed data and the background prediction, normalized by the total uncertainty.}
    \label{fig:postfit_VRs}
\end{figure}

The method performs well in tests with MC simulations.
To test the method with data, a validation region (VR), defined by inverting the sphericity requirement to $S < 0.1$ is used. 
The VR-FAIL and VR-PASS regions are once again separated by the SVM score of 0.63. 
The post-fit distributions of the \st variable in the VR-FAIL and VR-PASS regions are shown in the left and right plots of Fig.~\ref{fig:postfit_VRs}, respectively. 
The signal mass point B1, $\md = 2\TeV$, $\mbh = 10\TeV$, and $n = 2$ is also shown in the plots and indicates that there is negligible signal contamination. The pull distributions in the PASS region, namely the distributions of the ratios plotted in the lower panels, are found to be approximately symmetrical, with standard deviations consistent with unity and centered near zero.
The observed saturated log-likelihood $p$-value in the VR-PASS region is higher than 0.05, indicating good agreement between the background prediction and the data.
The higher level of agreement between the model and the data in the VR-FAIL region arises because this region is used to constrain the background estimate.
 
\section{Systematic uncertainties}
\label{sec:systematics}

The systematic uncertainties are incorporated as nuisance parameters and are profiled in the maximum likelihood fit.

The uncertainty in the integrated luminosity is applied as a 0.73\% uncertainty affecting the signal normalization~\cite{PAS-LUM-20-001}. The uncertainty associated with pileup is estimated by varying the total inelastic cross section~\cite{CMS:2018mlc}, which is used to compute the event weights, by one standard deviation.
This variation is then applied to the signal events to match the pileup distribution observed in data. 
The resulting variations  due to the pileup uncertainty are below 1\% across all signal mass points.

The uncertainties in the jet energy corrections~\cite{CMS:2020poo,CMS:2020ebo,CMS:2019ctu} are included in both background estimation methods. This leads to subdominant ($<1$\%) uncertainties.

For the SI background estimation method, the primary sources of systematic uncertainty are associated with extrapolating the background from the CR to the SR and its normalization.
The extrapolation uncertainty arises from the procedure used to model the background and it
accounts for the range of possible background functions that fit the data in the fit region.
The uncertainty is derived from the envelope of the highest and lowest fits obtained from the 16 test functions, which vary with multiplicity and \st.
For $\st > 5.2\TeV$, where the background prediction relies on an extrapolation beyond the fit region, this procedure results in a conservative uncertainty of approximately 50\% in the nominal background prediction for all values of $N$.

An additional uncertainty is introduced to account for residual variations and possible non-invariance in the low-\st region due to initial-state radiation. This uncertainty is assigned a baseline value of 30\% and increases with \st, reaching up to 120\% in the high-\st region.
A background normalization uncertainty associated with the procedure described above accounts for variations in the low-\st normalization region, and ranges from 0.6~to~10\%, depending on multiplicity.

For the PS method, the background uncertainty has two sources: the statistical uncertainty in the background yield in the FAIL region, which varies from 1\% to 100\% as \st increases, and the statistical uncertainty in \RPF, which varies from 1\% to 10\% depending on \st. 

Both methods are statistically limited.

\section{Results}
\label{sec:results}

The results of the search are extracted by performing a simultaneous maximum likelihood fit with the CMS statistical analysis tool \textsc{Combine}~\cite{CMS:2024onh}, which is based on the \textsc{RooFit}~\cite{Verkerke:2003ir} and \textsc{RooStats}~\cite{Moneta:2010pm} frameworks. They are interpreted in terms of model-independent limits on BSM physics in energetic, multiparticle final states, and as model-dependent limits for a set of semiclassical BH and SB scenarios, as well as for EW sphalerons.

Limits are set using the \CLs method~\cite{Junk:1999kv,Read:2002hq,CMS-NOTE-2011-005} with log-normal priors in the likelihood to constrain the nuisance parameters near their best fit values. We do not use the asymptotic approximation of the \CLs method~\cite{Cowan:2010js}, as most models in the search region have very low expected background,  a condition for which the asymptotic approximation is known to overestimate the search sensitivity.

\subsection{Model-independent results}

\begin{figure}[!htbp]
    \includegraphics[width=0.49\textwidth]{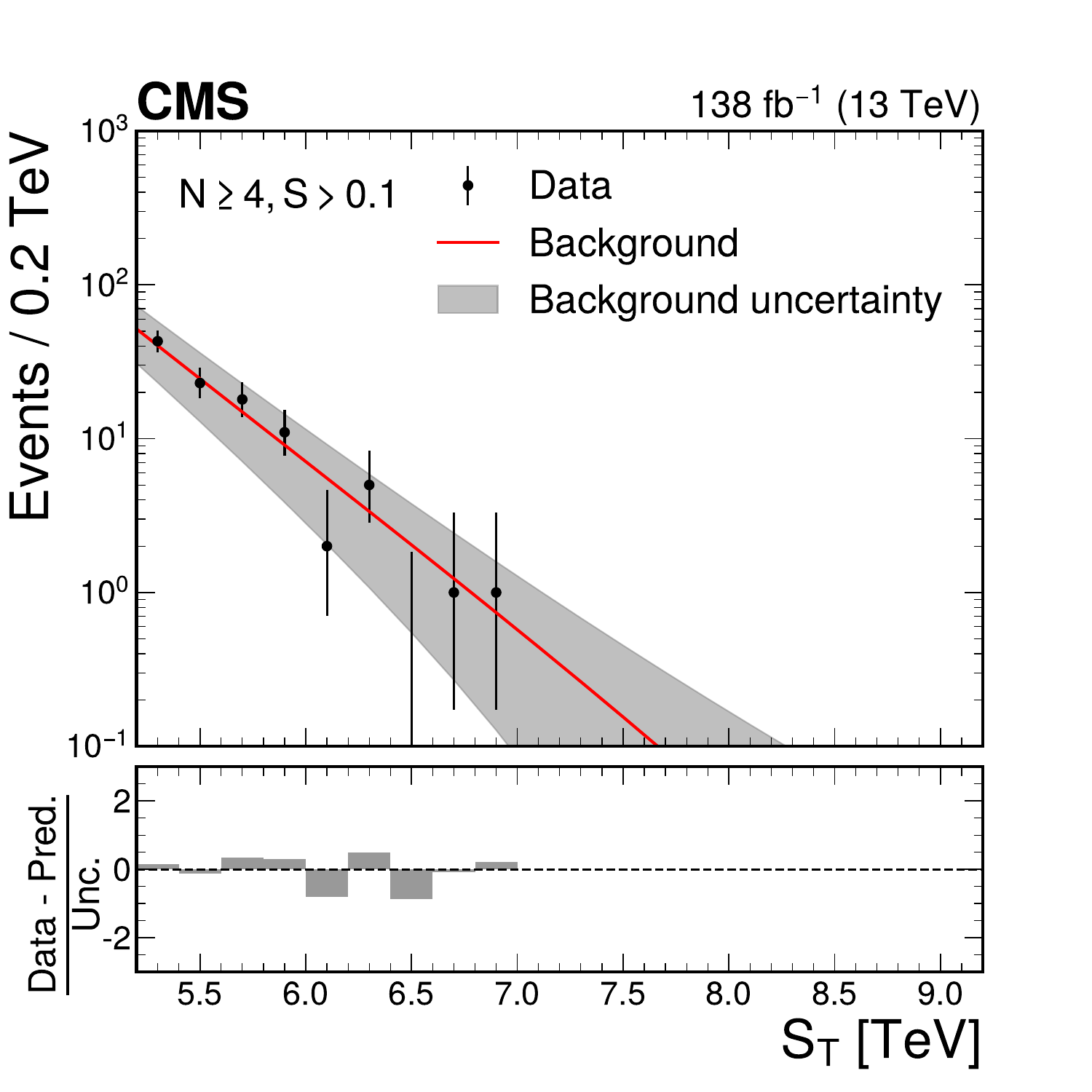}
    \includegraphics[width=0.49\textwidth]{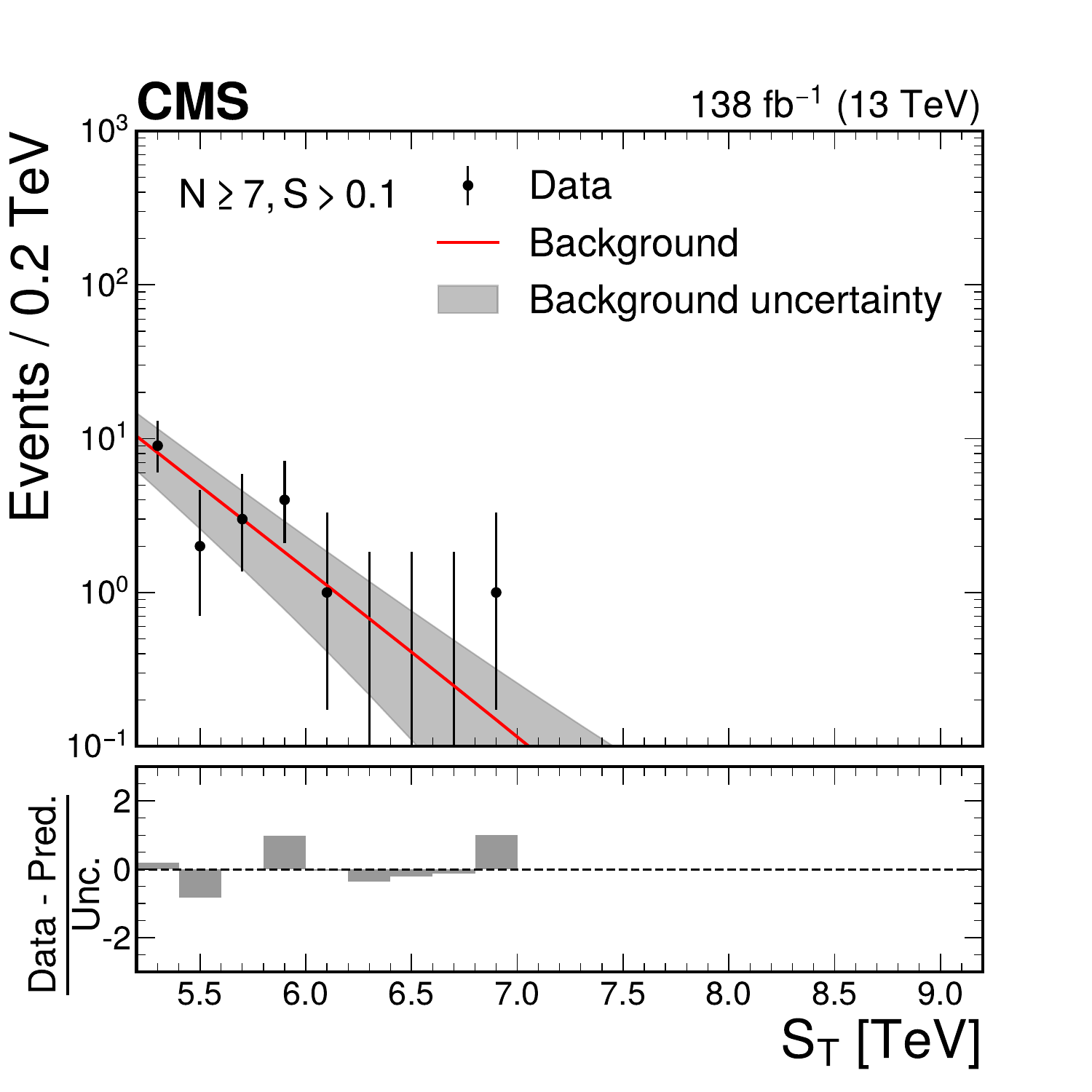}
    \caption{The \st distribution in the $N \geq 4$ (left) and $N \geq 7$ (right) SI-SR in data, indicated by the black dots, along with the background prediction and its uncertainty  represented by the red line and gray band, respectively. The error bars on the data points correspond to Poisson confidence intervals associated with the observed event counts. Lower panel as in Fig.~\ref{fig:postfit_SI_VR}.}
    \label{fig:traditional_st_prediction}
\end{figure}

Figure~\ref{fig:traditional_st_prediction} shows the \st distribution in the SI-SR in data along with the background prediction and its uncertainty.
No significant deviation in data compared to the expected SM background is observed.

Model-independent 95\% confidence level (\CL) limits on the signal cross section for $\st > \stmin$ times acceptance (A) are derived for a generic signal producing events above a threshold \stmin. These limits do not rely on assumptions about the signal model's properties or production mechanism.
Acceptance is defined with respect to a fiducial phase space corresponding to the full event and object selection used in the analysis. This includes the requirements on reconstructed objects, event multiplicity, and the sphericity selection $S > 0.1$.
For sphaleron signal models, the acceptance is typically high, ranging from 89\% to 93\%. For microscopic black hole signals, the acceptance depends strongly on the mass point and model parameters. In the case of the B1 model, the acceptance varies between approximately 50\% and 90\% across the considered parameter space.
The observed number of events in the SI-SR sample is compared to the expected number of background events coming from the extrapolation of the fitted function, 
allowing for a limit to be set on any excess of events that could indicate the presence of a signal.

\begin{figure}[!htbp]
    \includegraphics[width=0.49\textwidth]{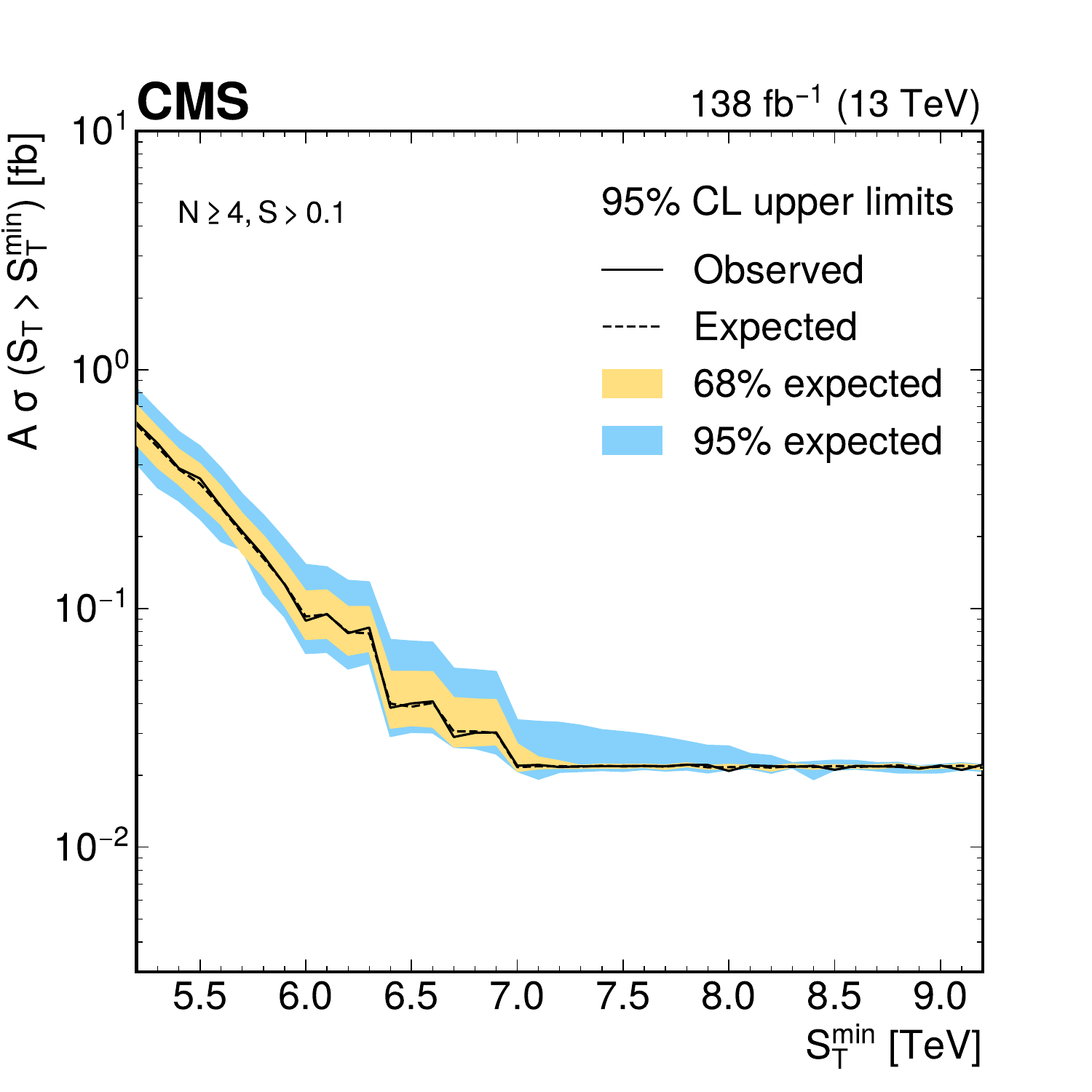}
    \includegraphics[width=0.49\textwidth]{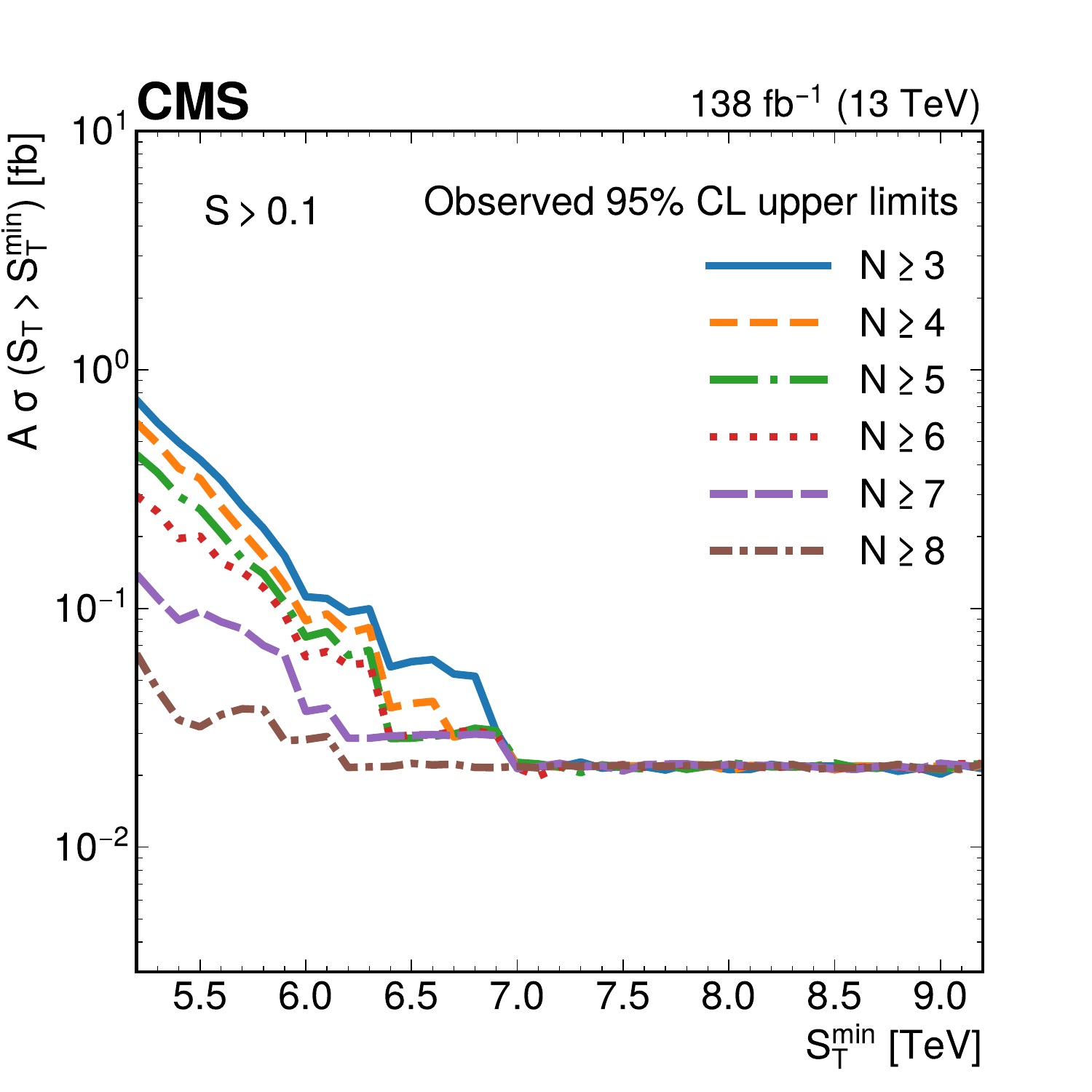}
    \caption{Expected and observed model-independent 95\% \CL upper limits on the cross section times acceptance for multiplicity $N \geq 4$, where the inner (outer) band represents the 68\% (95\%) quantile of the expected limit (left), and the observed limits with different minimum object multiplicity requirements (right).}
    \label{fig:MIlimits_expband_sphericity}
\end{figure}

Figure~\ref{fig:MIlimits_expband_sphericity} (left) shows the expected median limits, with 68\% and 95\% quantile bands, and observed limits for $N \geq 4$.
We show the observed limits for different object multiplicity thresholds in Fig.~\ref{fig:MIlimits_expband_sphericity} (right) as functions of the \st threshold. The observed limits show an improvement of about a factor of 4 over the previous CMS analysis~\cite{CMS:2018ozv}. The gain comes from both the increased integrated luminosity and the enhanced background rejection provided by the sphericity selection. 

\subsection{Model-dependent results}

\begin{figure}[ht!]
    \centering
     \includegraphics[width=0.98\textwidth]{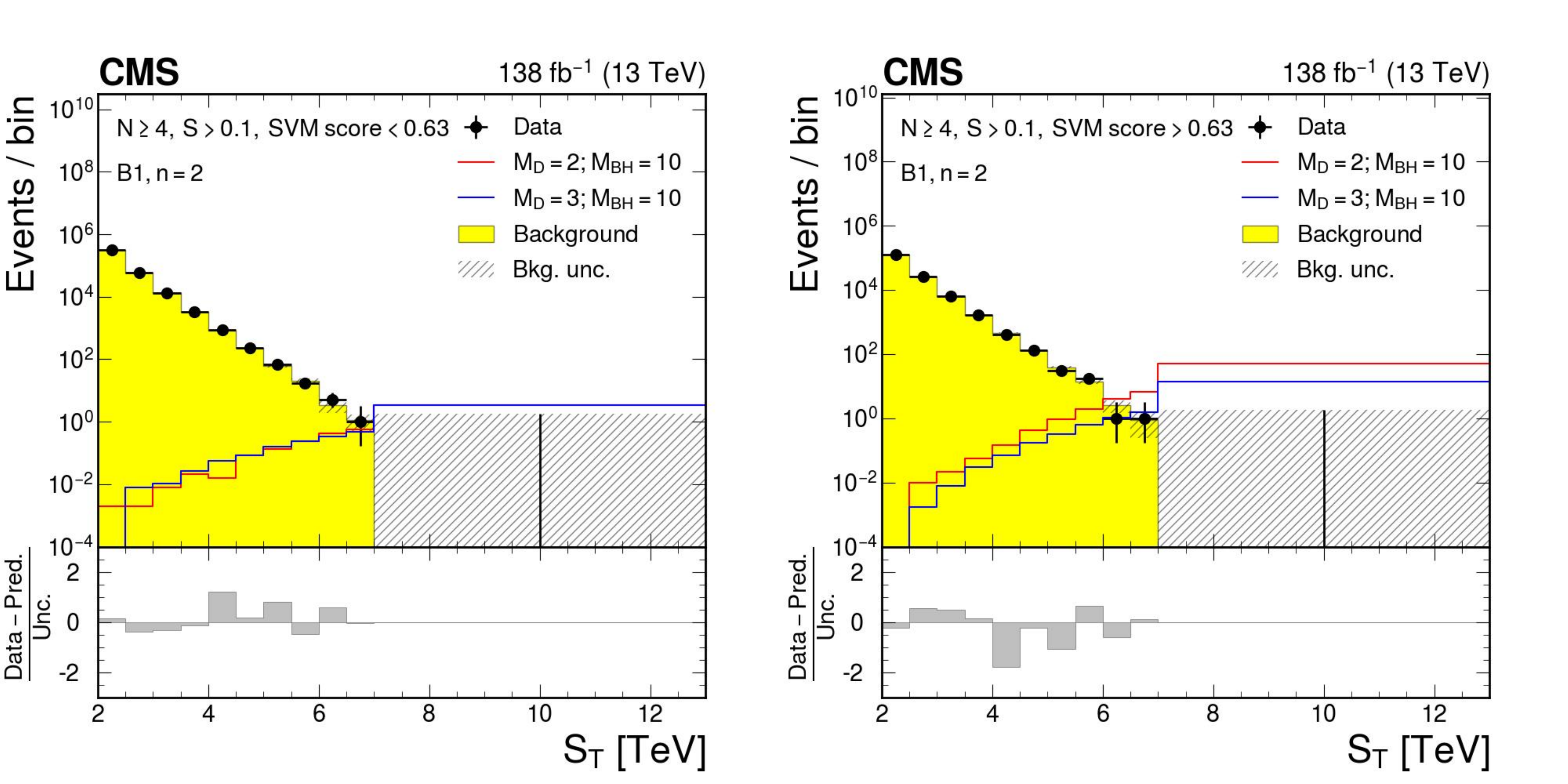}
    \caption{Post-fit \st distributions in the FAIL (left) and PASS (right) regions in data. The gray shaded area includes both statistical and systematic uncertainties in the background prediction (yellow histogram) while the red and blue lines are B1 signal examples, as noted in the legends. Lower panel as in Fig.~\ref{fig:postfit_VRs}.}
    \label{fig:pass_fail_SR}
\end{figure}

Model-dependent results are derived using the PS method for background estimation, with limits set using the \st distribution in the SR, defined by $N \geq 4$, $S > 0.1$, and SVM score $>0.63$.
The background prediction and observed data for the PS method are shown in Fig.~\ref{fig:pass_fail_SR}, together with the B1 model at $n = 2$ with $\md = 2\TeV$, $\mbh = 10\TeV$ (red) and with $\md = 3\TeV$, $\mbh = 10\TeV$ (blue) as examples of expected signal distributions. No significant deviation in data compared to the expected SM background is observed.

\begin{figure}[ht!]
    \centering
    \includegraphics[width=0.49\textwidth]{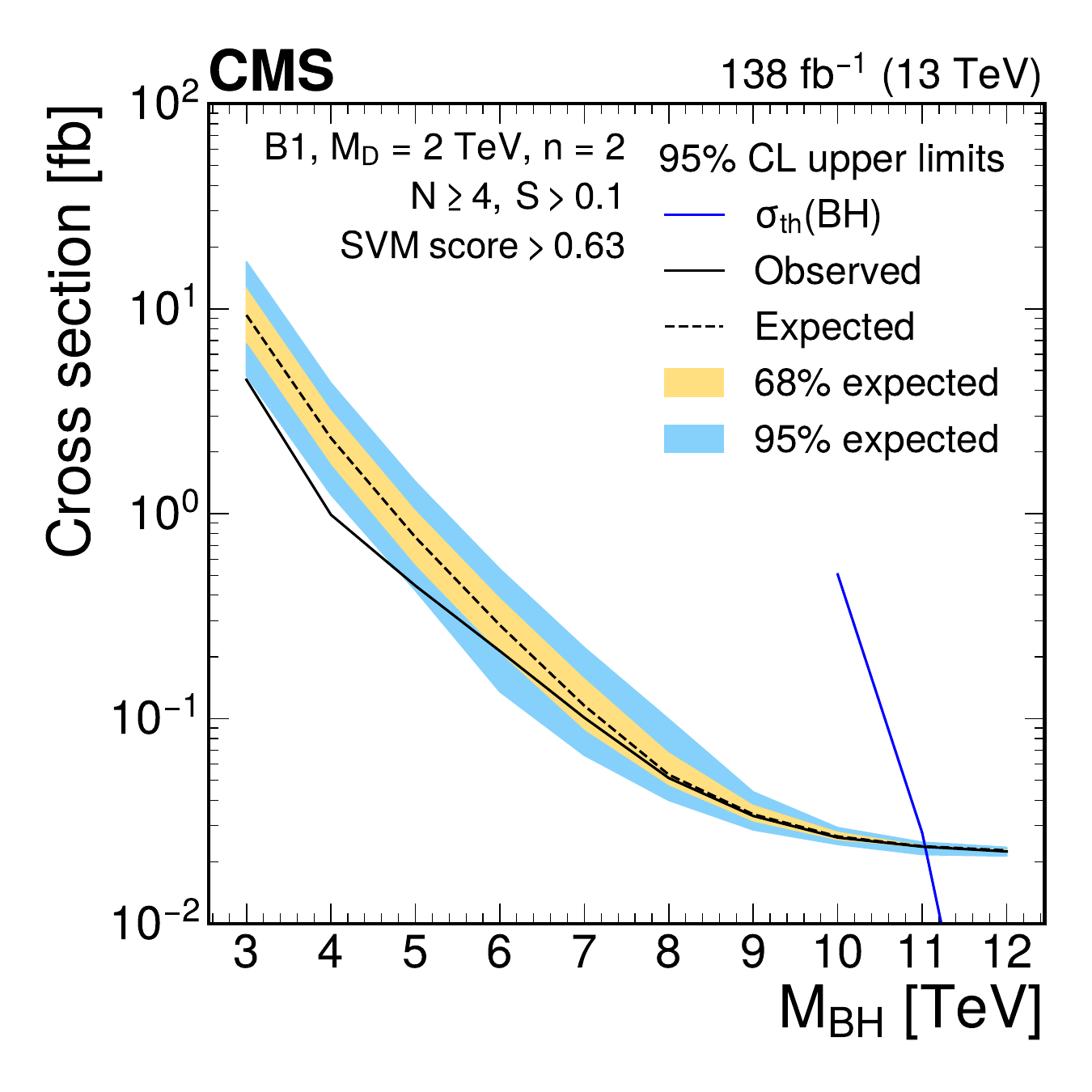}
    \includegraphics[width=0.49\textwidth]{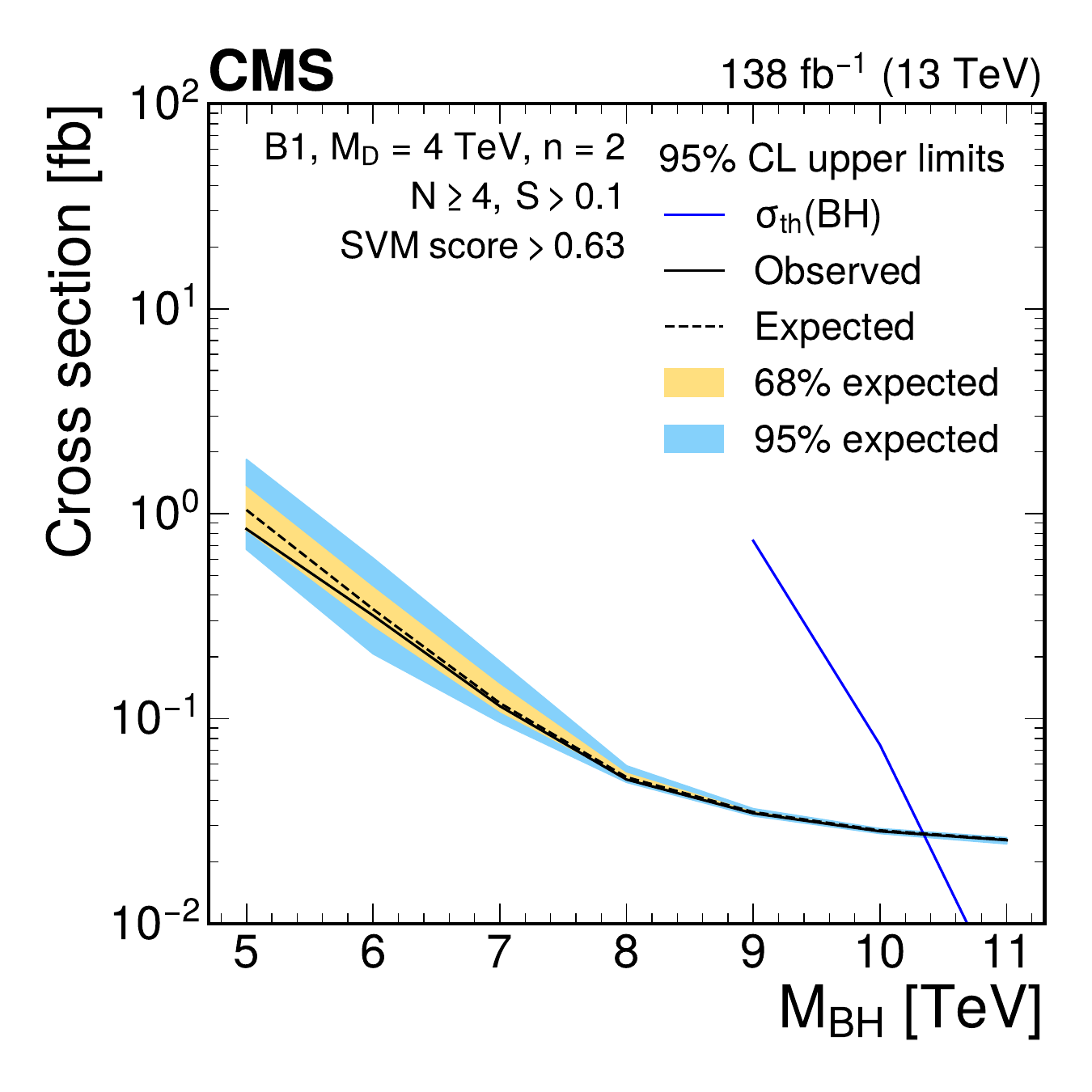}
    \caption{Expected and observed 95\% \CL upper limits on the cross section for a semiclassical nonrotating BH model (B1) with $n = 2$ and $\md = 2\TeV$ (left) or  $\md = 4\TeV$ (right), as a function of \mbh. The blue curves represent the theoretical cross section values. The inner (outer) band represents the 68\% (95\%) quantile of the expected limit.}
    \label{fig:B1_MD2_MD4}
\end{figure}

In Fig.~\ref{fig:B1_MD2_MD4}, we compare the observed upper limits on the signal cross section to the predicted cross section. In the B1 model with $n = 2$, values of \mbh below 11.0\TeV are excluded for $\md = 2\TeV$, and values below 10.4\TeV are excluded for $\md = 4\TeV$.

The observed limit lies below the median expected limit at lower \mbh values because the observed yield in the PASS region around 4\TeV in \st, visible in Fig.~\ref{fig:pass_fail_SR}, is below the predicted background. While this bin contributes negligibly to the limits at high \mbh, signal models with lower threshold masses populate this region of \st. The 68\% and 95\% expected uncertainty bands narrow with increasing \mbh as the background prediction becomes very small causing the expected limits to approach the background-free regime where the exclusion is determined primarily by the signal yield required for a 95\% \CL exclusion. However, the bands retain finite width across the full \mbh range.

\begin{figure}[ht!]
    \centering
    \includegraphics[width=0.49\textwidth]{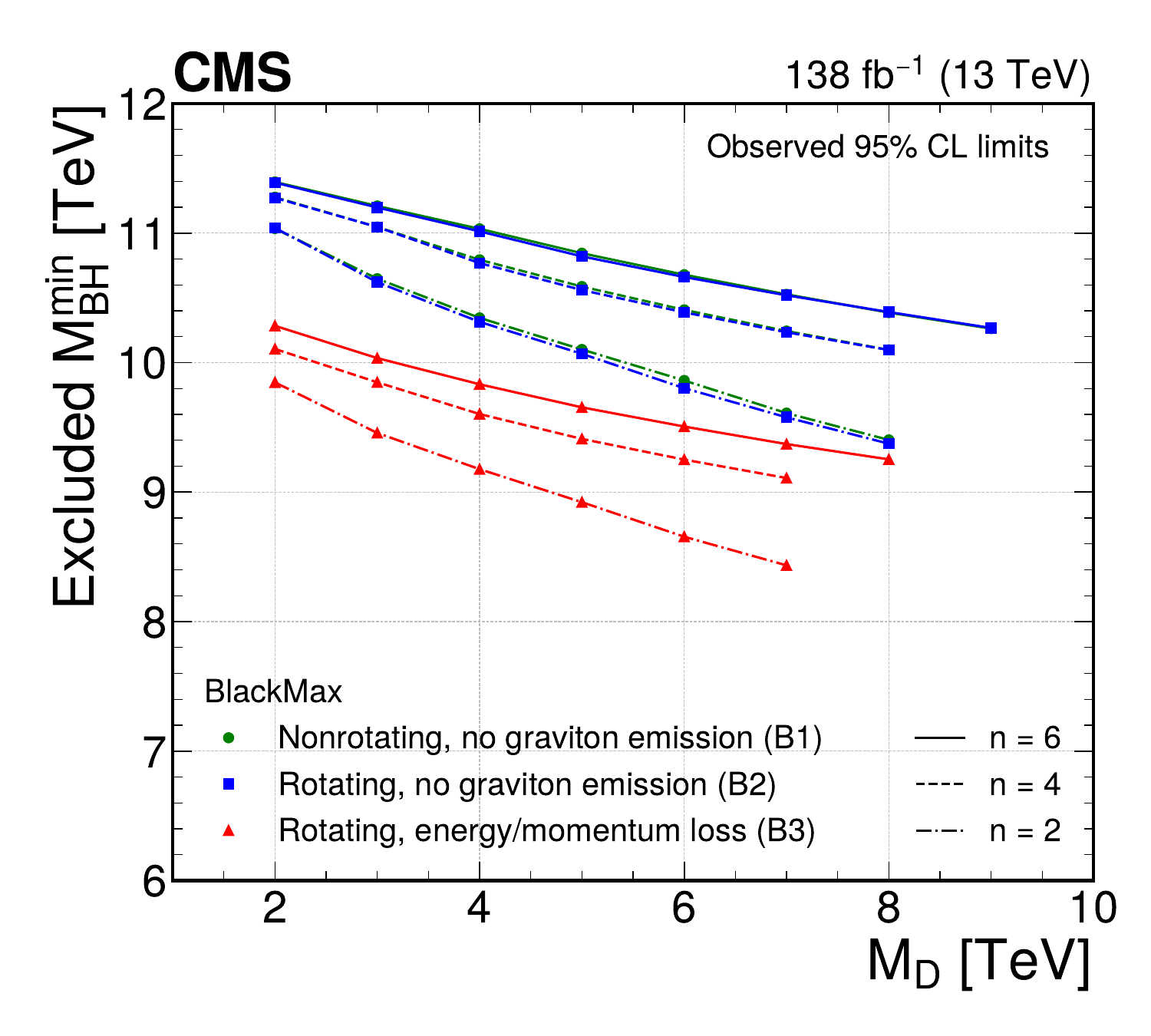}
    \includegraphics[width=0.49\textwidth]{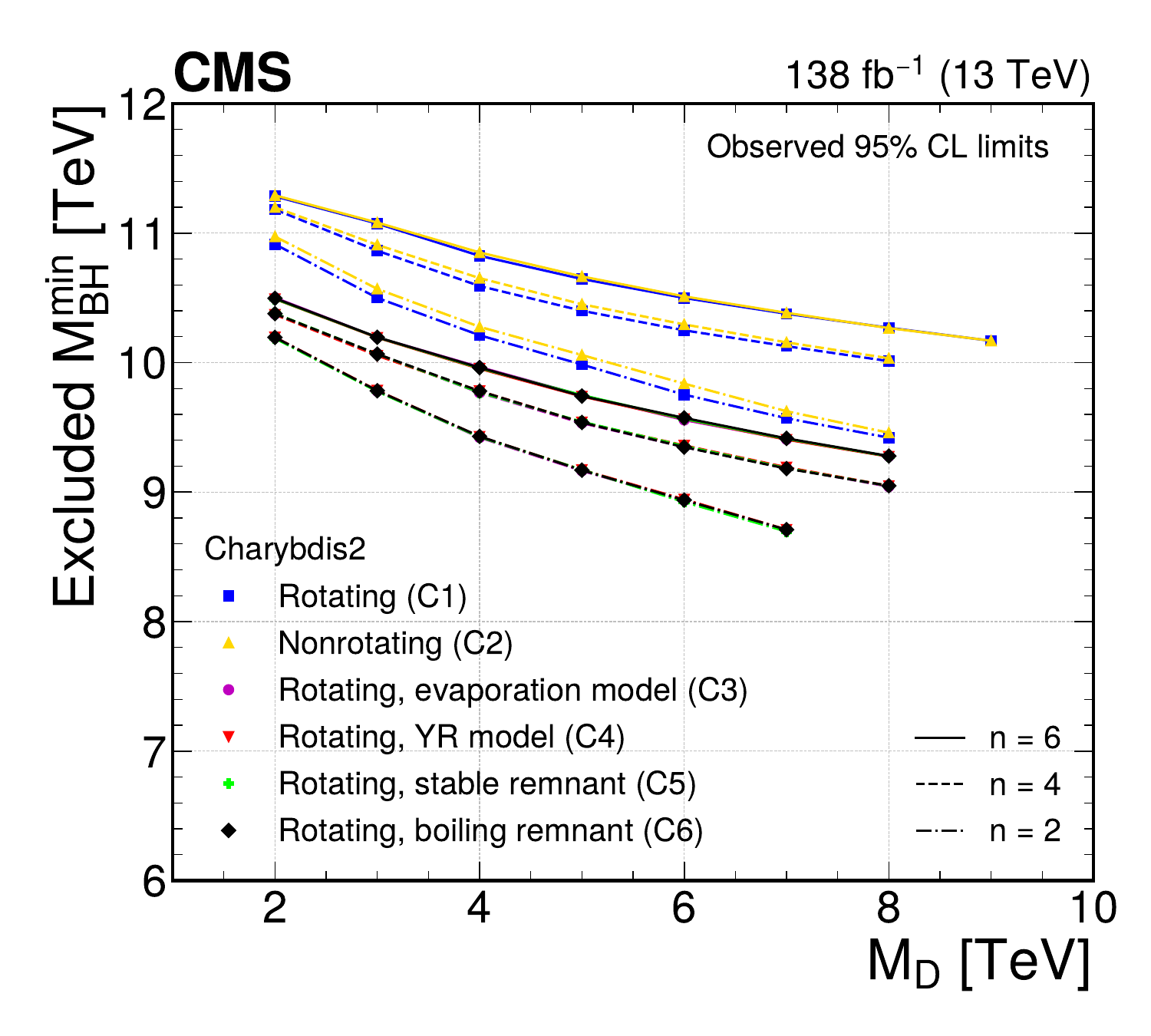}
    \caption{Excluded \mbhmin values as functions of \md and $n$ for a variety of \textsc{BlackMax} (left) and \textsc{Charybdis2} (right) BH models.}
    \label{fig:mass_limit_blackhole}
\end{figure}

Figure~\ref{fig:mass_limit_blackhole} summarizes results for excluded masses as a function of \md for a variety of BH models, and as functions of the number of extra dimensions $n$. 
The excluded \mbhmin decreases with increasing \md and increases with $n$, as expected from the corresponding behavior of the BH production cross section in the ADD framework. 
Rotating BH scenarios with full evaporation (B2, C1, C2) yield the strongest limits, while models with energy and momentum loss due to graviton emission (B3) or with the remnant formation (C5, C6) give weaker exclusions due to reduced visible final-state energy. 
Overall, \mbhmin values between approximately 9 and 11.5\TeV are excluded at 95\% CL across the considered parameter space.

Taking the B1 model with $\md = 4\TeV$ and $n = 4$ as an example, the excluded BH mass value is extended by about 1.4\TeV with respect to the previous CMS search, where 20\% of the improvement is a result of increased integrated luminosity, 20\% is because of enhanced selections resulting in better signal-to-background discrimination, and 60\% comes from updating PDFs from MSTW2008LO to the recent NNPDF3.1 set.
This dominant PDF-driven gain is expected, as the sensitivity of the analysis is governed by events probing large parton momentum fractions, where updated PDF parameterizations lead to sizable changes in the predicted signal yield.

An alternative interpretation of these results is obtained by establishing the maximum allowed number of extra dimensions as a function of the fundamental Planck scale mass \md and the minimum black hole mass \mbhmin for different BH models.
Upper limits on the number of extra dimensions \nmax are derived as the value of $n$ at the intersection between the experimental limit curves and the theoretical cross sections, floored to the nearest integer. This limit is well-defined because the theoretical cross section is a monotonically increasing function of $n$. When no intersection is found within the studied range ($n\in[2,6]$), the result is reported as $\nmax > 6$ if the experimental limit lies above all theoretical cross sections, indicating that the exclusion does not reach $n = 6$, or as $\nmax < 2$ if it lies below, indicating that the exclusion reaches beyond $n = 2$.

At 95\% \CL, more than two extra dimensions are excluded under different assumptions of \md and \mbhmin for the majority of the models considered. A systematic scan over both \md and \mbhmin values is performed and the resulting parameter combinations that exclude a given number of extra dimensions are presented in Table~\ref{tab:nextra_limits}.
For example, for models B1 and B2, the combinations $\md \in (2-8)$\TeV with $\mbhmin \in (3,9)\TeV$, $(2-5)$\TeV with $\mbhmin = 10$\,\TeV, and $\md = 2\TeV$ with $\mbhmin = 11$\TeV all yield $\nmax < 2$, \ie, the allowed number of extra dimensions are zero or one. The last row shows the combinations, typically the highest $\mbhmin$ with high $\md$, when we cannot claim any exclusions. Combinations with $\mbhmin < \md$ are unphysical and not considered.
A result of $\nmax < 2$ implies that, for the corresponding ($\md$, $\mbhmin$) combination and generator, black hole production is excluded for all values of $n$ that are viable in the ADD model.

\begin{table}[ht!]
\centering
\topcaption{Parameter combinations of the fundamental Planck scale \md and minimum black hole mass \mbhmin (both in \TeV) that exclude a given number of extra dimensions \nmax, for BH models generated with \textsc{BlackMax} and \textsc{Charybdis2}. Each row collects all ($\md$, $\mbhmin$) pairs yielding the same exclusion level. Combinations with $\mbhmin < \md$ are unphysical and not considered. The first row indicates the pairs that are excluded for any number of extra dimensions in the ADD model, while the last row shows the parameters where no exclusion could be made in this analysis.}
\label{tab:nextra_limits}
\renewcommand{\arraystretch}{1.15}
\begin{tabular}{cccccc}
\hline\noalign{\smallskip}
\multicolumn{5}{c}{($\md$\,[\TeVns],\ $\mbhmin$\,[\TeVns])} & \\
\noalign{\smallskip}\cline{1-5}\noalign{\smallskip}
{B1--B2} & {B3} & {C1} & {C2} & {C3--C6} & {\nmax} \\
\noalign{\smallskip}\hline\noalign{\smallskip}

  \begin{tabular}[c]{@{}c@{}}(2--8, 3--9)\\ (2--5, 10)\\ (2, 11)\end{tabular}
  & \begin{tabular}[c]{@{}c@{}}(2--7, 3--9)\\ (2--4, 9)\\ (2, 11)\end{tabular}
  & \begin{tabular}[c]{@{}c@{}}(2--8, 3--9)\\ (2--4, 10)\end{tabular}
  & \begin{tabular}[c]{@{}c@{}}(2--8, 3--9)\\ (2--5, 10)\end{tabular}
  & \begin{tabular}[c]{@{}c@{}}(2--7, 3--9)\\ (2--5, 9)\\ (2, 10)\end{tabular}
  & $<$2 \\
\noalign{\smallskip}\hline\noalign{\smallskip}

  \begin{tabular}[c]{@{}c@{}}(6,10)\\ (7, 10)\end{tabular}
  & \begin{tabular}[c]{@{}c@{}}(5--6,9)\\ (2, 10)\end{tabular}
  & \begin{tabular}[c]{@{}c@{}}(5--6,10)\\ (2, 11)\end{tabular}
  & \begin{tabular}[c]{@{}c@{}}(6,10)\\ (2, 11)\end{tabular}
  & \begin{tabular}[c]{@{}c@{}}(6--7,9)\end{tabular}
  & 2 \\
\noalign{\smallskip}\hline\noalign{\smallskip}

  \begin{tabular}[c]{@{}c@{}}(8,10)\\ (3, 11)\end{tabular}
  & \begin{tabular}[c]{@{}c@{}}(7,9)\end{tabular}
  & \begin{tabular}[c]{@{}c@{}}(7--8,10)\end{tabular}
  & \begin{tabular}[c]{@{}c@{}}(7--8,10)\end{tabular}
  & \begin{tabular}[c]{@{}c@{}}(8,9)\\ (3,10)\end{tabular}
  & 3 \\
\noalign{\smallskip}\hline\noalign{\smallskip}

  \begin{tabular}[c]{@{}c@{}}(9,10)\end{tabular}
  & \begin{tabular}[c]{@{}c@{}}(8,9)\end{tabular}
  & \begin{tabular}[c]{@{}c@{}}(9,10)\end{tabular}
  & \begin{tabular}[c]{@{}c@{}}(9,10)\\ (3,11)\end{tabular}
  & \begin{tabular}[c]{@{}c@{}}\NA\end{tabular}
  & 4 \\
\noalign{\smallskip}\hline\noalign{\smallskip}

  \begin{tabular}[c]{@{}c@{}}(4,11)\end{tabular}
  & \begin{tabular}[c]{@{}c@{}}(3,10)\end{tabular}
  & \begin{tabular}[c]{@{}c@{}}(3,11)\end{tabular}
  & \begin{tabular}[c]{@{}c@{}}\NA\end{tabular}
  & \begin{tabular}[c]{@{}c@{}}\NA\end{tabular}
  & 5 \\
\noalign{\smallskip}\hline\noalign{\smallskip}

  \begin{tabular}[c]{@{}c@{}}(5--9,11)\end{tabular}
  & \begin{tabular}[c]{@{}c@{}}(4--9,10)\\ (2--9,11)\end{tabular}
  & \begin{tabular}[c]{@{}c@{}}(4--9,11)\end{tabular}
  & \begin{tabular}[c]{@{}c@{}}(4--9,11)\end{tabular}
  & \begin{tabular}[c]{@{}c@{}}(4--9,10)\\ (2--9,11)\end{tabular}
  & $>$6 \\

\noalign{\smallskip}\hline

\end{tabular}
\end{table}

\begin{figure}[ht!]
    \centering
    \includegraphics[width=0.49\textwidth]{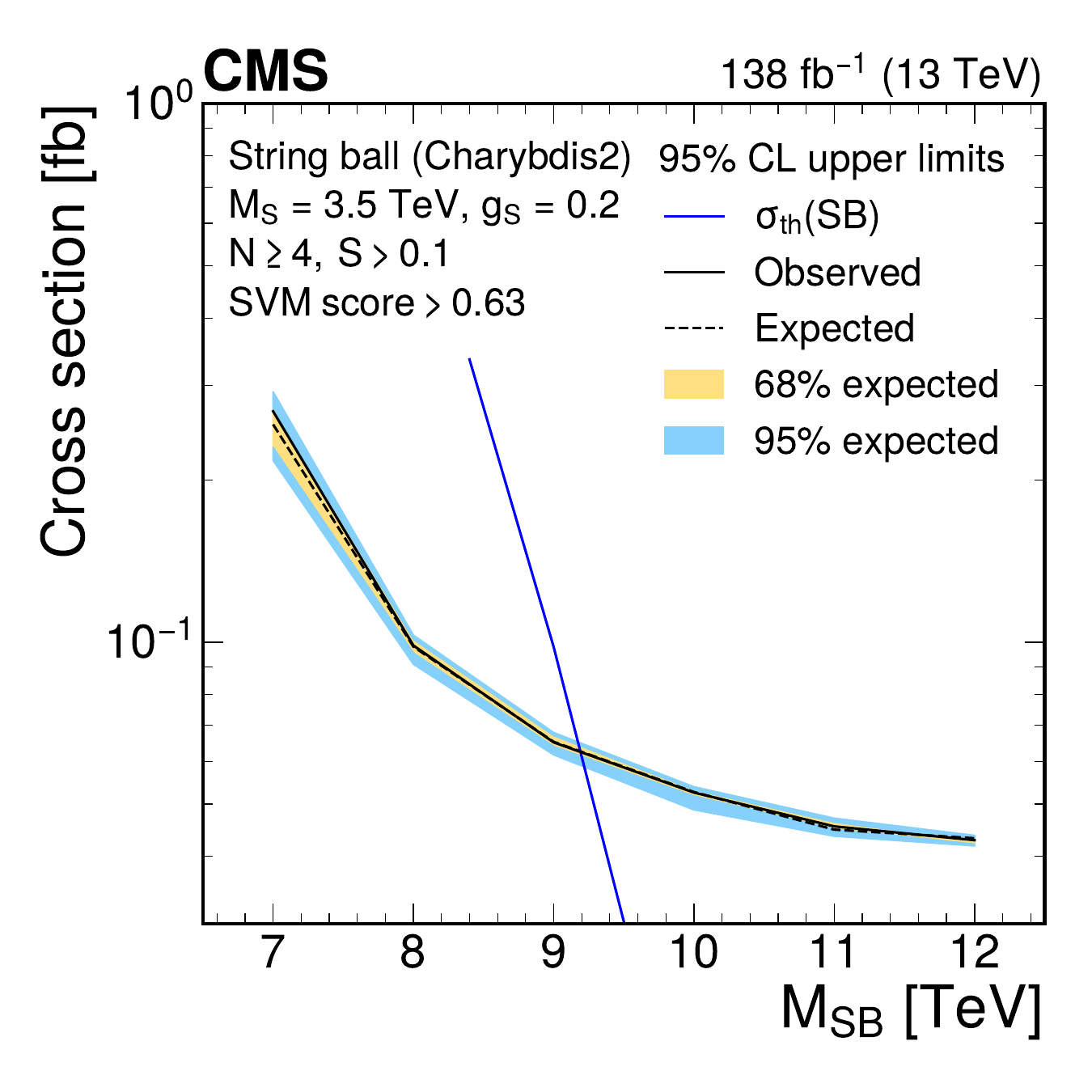}
    \includegraphics[width=0.49\textwidth]{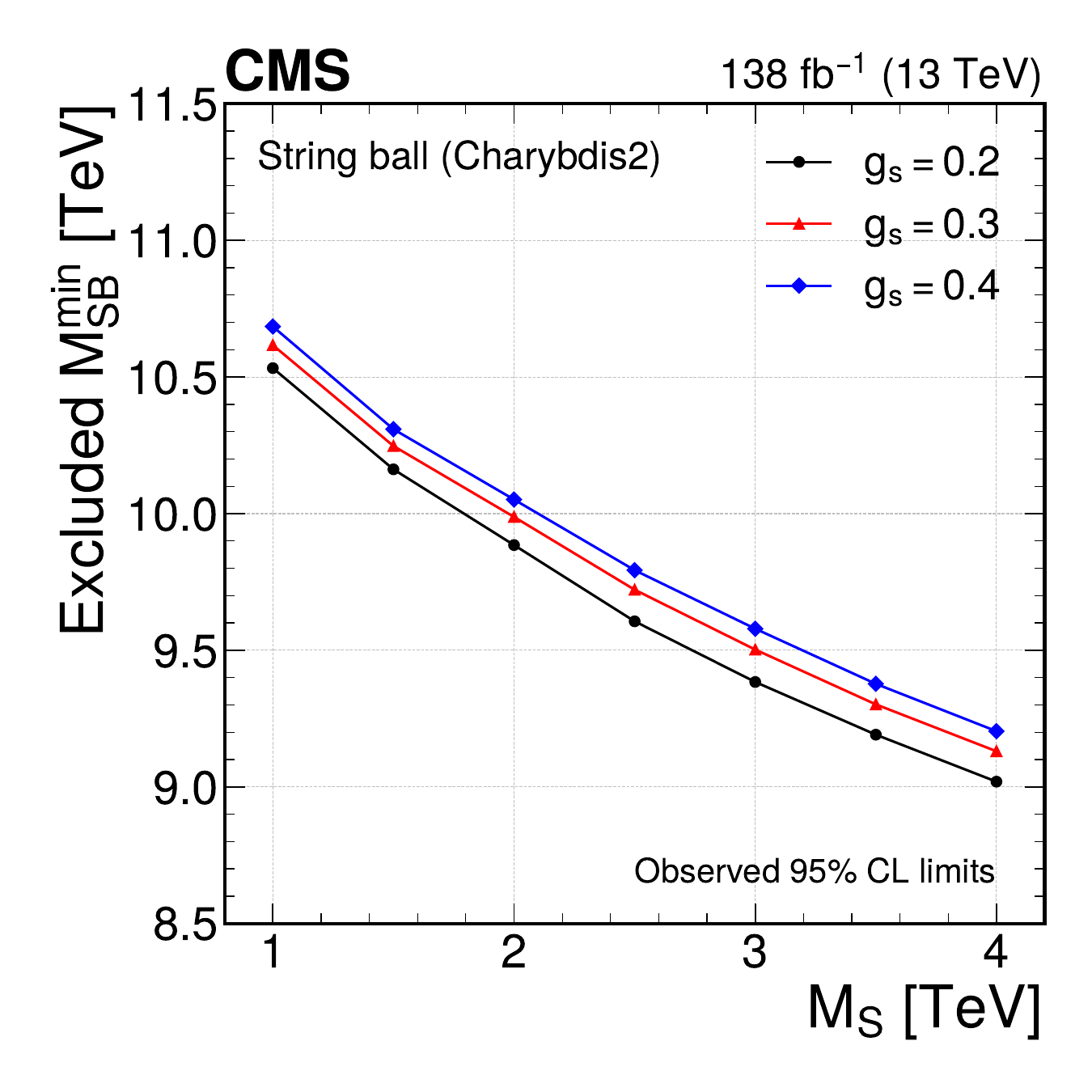}
    \caption{Expected and observed 95\% \CL upper limits for SB models with $\ms = 3.5\TeV$ and $\gs = 0.2$ (left) and excluded SB mass values as functions of \ms at $\gs=0.2, 0.3$, and $0.4$ (right). The inner (outer) band represents the 68\% (95\%) quantile of the expected limit.}
    \label{fig:limit_stringball}
\end{figure}

The upper limits on the SB production cross sections as functions of the SB mass are set for different \ms and \gs values. An example for $\ms = 3.5\TeV$ and $\gs = 0.2$ is shown in Fig.~\ref{fig:limit_stringball} (left). The excluded masses of the SB as functions of \ms at different \gs values are summarized in Fig.~\ref{fig:limit_stringball} (right).

\begin{figure}[ht!]
    \centering
    \includegraphics[width=0.49\textwidth]{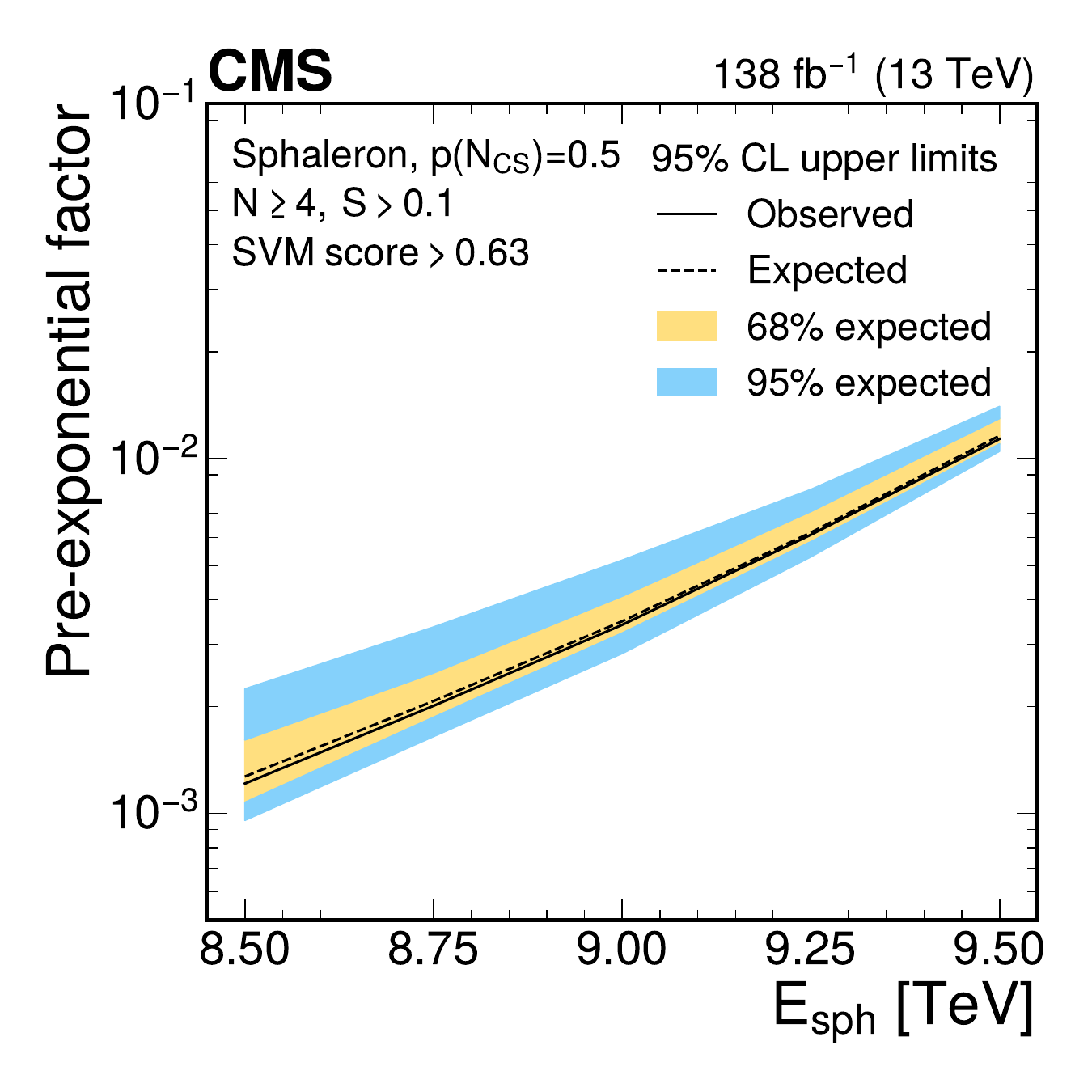}
    \includegraphics[width=0.49\textwidth]{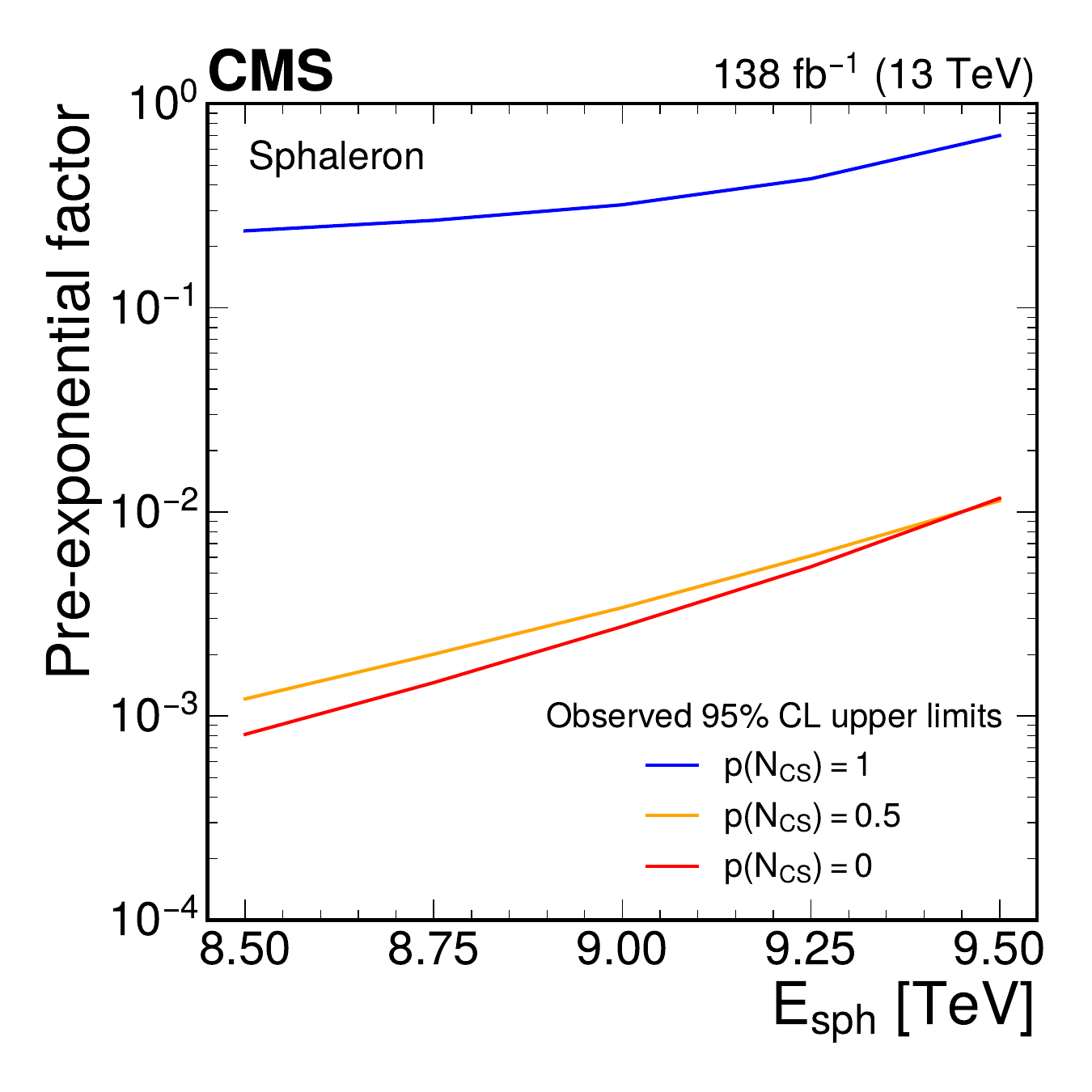}
    \caption{Expected and observed 95\% \CL upper limits on the pre-exponential factor for the sphaleron model with $\pncs=0.5$ (left), and observed limits with $\pncs=0$, $0.5$, and $1$ (right). The inner (outer) band (left) represents the 68\% (95\%) quantiles of the expected limit.}
    \label{fig:sphaleron_limit_PEF}
\end{figure}

For the sphaleron signal, a limit on the cross section can be converted into a limit on the pre-exponential factor, defined in Section~\ref{sec:signals}. Following the method in Ref.~\cite{Ellis:2016ast}, we calculate the limits on the pre-exponential factor for five scanned \esph values: 8.5, 8.75, 9, 9.25, 9.5\TeV, under the assumptions of $\pncs = 0$, $0.5$, and $1$. The observed and expected 95\% \CL upper limits on the pre-exponential factor as functions of \esph are shown in Fig.~\ref{fig:sphaleron_limit_PEF}.
The observed (expected) upper limit for the nominal $\esph = 9\TeV$ and $\pncs = 0.5$ is 0.0034 (0.0035), which is approximately 6.2 (3.4) times more stringent than the previous best limit of 0.021 (0.012) from CMS~\cite{CMS:2018ozv}.

\newpage
\section{Summary}
\label{sec:summary}

A dedicated search for black holes, string balls, and sphalerons produced in proton-proton collisions at $\sqrt{s} = 13\TeV$ using data collected with the CMS detector has been presented.

No excesses above the standard model background predictions are observed. 
We set 95\% confidence level (CL) model-independent limits on cross section of hypothetical signals characterized by a large multiplicity of energetic jets, leptons, and photons.
The model-independent results demonstrate approximately a factor of four improvement in the cross section limit compared to the previous CMS analysis.
The model-dependent results exclude at 95\% CL semiclassical black holes and string balls with masses below 8.4--11.4\TeV and 9.0--10.7\TeV, respectively, depending on the model and the number of extra dimensions. This extends the exclusion reach by 1--1.6\TeV and 1.3--1.9\TeV, respectively.
The observed (expected) upper limit on the sphaleron pre-exponential factor for the nominal electroweak sphaleron transition energy of 9\TeV is 0.0034 (0.0035) at 95\% CL, which is strengthened by a factor of 6.2 (3.4) compared to the previous best limit of 0.021 (0.012) from CMS~\cite{CMS:2018ozv}.
These are the most stringent limits on the sphaleron pre-exponential factor to date.
A significant improvement in the model-dependent study over previous results comes from an improved understanding of parton distribution functions. Additional significant gains can be traced to both the increased integrated luminosity, and the enhanced background rejection provided by the sphericity and phase space distance event selection requirements.

\begin{acknowledgments}
We congratulate our colleagues in the CERN accelerator departments for the excellent performance of the LHC and thank the technical and administrative staffs at CERN and at other CMS institutes for their contributions to the success of the CMS effort. In addition, we gratefully acknowledge the computing centers and personnel of the Worldwide LHC Computing Grid and other centers for delivering so effectively the computing infrastructure essential to our analyses. Finally, we acknowledge the enduring support for the construction and operation of the LHC, the CMS detector, and the supporting computing infrastructure provided by the following funding agencies: SC (Armenia), BMBWF and FWF (Austria); FNRS and FWO (Belgium); CNPq, CAPES, FAPERJ, FAPERGS, and FAPESP (Brazil); MES and BNSF (Bulgaria); CERN; CAS, MoST, and NSFC (China); MINCIENCIAS (Colombia); MSES and CSF (Croatia); RIF (Cyprus); SENESCYT (Ecuador); ERC PRG and PSG, TARISTU24-TK10 and MoER TK202 (Estonia); Academy of Finland, MEC, and HIP (Finland); CEA and CNRS/IN2P3 (France); SRNSF (Georgia); BMFTR, DFG, and HGF (Germany); GSRI (Greece); MATE and NKFIH (Hungary); DAE and DST (India); IPM (Iran); SFI (Ireland); INFN (Italy); MSIT and NRF (Republic of Korea); MES (Latvia); LMTLT (Lithuania); MOE and UM (Malaysia); BUAP, CINVESTAV, CONACYT, LNS, SEP, and UASLP-FAI (Mexico); MOS (Montenegro); MBIE (New Zealand); PAEC (Pakistan); MSHE, NSC, and NAWA (Poland); FCT (Portugal); MESTD (Serbia); MICIU/AEI and PCTI (Spain); MOSTR (Sri Lanka); Swiss Funding Agencies (Switzerland); MST (Taipei); MHESI (Thailand); TUBITAK and TENMAK (T\"{u}rkiye); NASU (Ukraine); STFC (United Kingdom); DOE and NSF (USA).

\hyphenation{Rachada-pisek} Individuals have received support from the Marie-Curie program and the European Research Council and Horizon 2020 Grant, contract Nos.\ 675440, 724704, 752730, 758316, 765710, 824093, 101115353, 101002207, 101001205, and COST Action CA16108 (European Union); the Leventis Foundation; the Alfred P.\ Sloan Foundation; the Alexander von Humboldt Foundation; the Science Committee, project no. 22rl-037 (Armenia); the Fonds pour la Formation \`a la Recherche dans l'Industrie et dans l'Agriculture (FRIA) and Fonds voor Wetenschappelijk Onderzoek contract No. 1228724N (Belgium); the Beijing Municipal Science \& Technology Commission, No. Z191100007219010, the Fundamental Research Funds for the Central Universities, the Ministry of Science and Technology of China under Grant No. 2023YFA1605804, the Natural Science Foundation of China under Grant No. 12535004, and USTC Research Funds of the Double First-Class Initiative No.\ YD2030002017 (China); the Ministry of Education, Youth and Sports (MEYS) of the Czech Republic; the Shota Rustaveli National Science Foundation, grant FR-22-985 (Georgia); the Deutsche Forschungsgemeinschaft (DFG), among others, under Germany's Excellence Strategy -- EXC 2121 ``Quantum Universe" -- 390833306, and under project number 400140256 - GRK2497; the Hellenic Foundation for Research and Innovation (HFRI), Project Number 2288 (Greece); the Hungarian Academy of Sciences, the New National Excellence Program - \'UNKP, the NKFIH research grants K 131991, K 138136, K 143460, K 143477, K 147557, K 146913, K 146914, K 147048, TKP2021-NKTA-64, and 2025-1.1.5-NEMZ\_KI-2025-00004, and MATE KKP and KKPCs Research Excellence and Flagship Research Groups grants (Hungary); the Council of Science and Industrial Research, India; ICSC -- National Research Center for High Performance Computing, Big Data and Quantum Computing, FAIR -- Future Artificial Intelligence Research, and CUP I53D23001070006 (Mission 4 Component 1), funded by the NextGenerationEU program, the Italian Ministry of University and Research (MUR) under Bando PRIN 2022 -- CUP I53C24002390006, PRIN PRIMULA 2022RBYK7T (Italy); the Latvian Council of Science; the Ministry of Science and Higher Education, project no. 2022/WK/14, and the National Science Center, contracts Opus 2021/41/B/ST2/01369, 2021/43/B/ST2/01552, 2023/49/B/ST2/03273, and the NAWA contract BPN/PPO/2021/1/00011 (Poland); the Funda\c{c}\~ao para a Ci\^encia e a Tecnologia (Portugal); the National Priorities Research Program by Qatar National Research Fund; MICIU/AEI/10.13039/501100011033, ERDF/EU, ``European Union NextGenerationEU/PRTR", projects PID2022-142604OB-C21, PID2022-139519OB-C21, PID2023-147706NB-I00, PID2023-148896NB-I00, PID2023-146983NB-I00, PID2023-147115NB-I00, PID2023-148418NB-C41, PID2023-148418NB-C42, PID2023-148418NB-C43, PID2023-148418NB-C44, PID2024-158190NB-C22, RYC2021-033305-I, RYC2024-048719-I, CNS2023-144781, CNS2024-154769 and Plan de Ciencia, Tecnolog{\'i}a e Innovaci{\'o}n de Asturias, Spain; the Chulalongkorn Academic into Its 2nd Century Project Advancement Project, the National Science, Research and Innovation Fund program IND\_FF\_68\_369\_2300\_097, and the Program Management Unit for Human Resources \& Institutional Development, Research and Innovation, grant B39G680009 (Thailand); the Eric \& Wendy Schmidt Fund for Strategic Innovation through the CERN Next Generation Triggers project under grant agreement number SIF-2023-004; the Kavli Foundation; the Nvidia Corporation; the SuperMicro Corporation; the Welch Foundation, contract C-1845; and the Weston Havens Foundation (USA).    
\end{acknowledgments}

\bibliography{auto_generated} 

\cleardoublepage \appendix\section{The CMS Collaboration \label{app:collab}}\begin{sloppypar}\hyphenpenalty=5000\widowpenalty=500\clubpenalty=5000\input{EXO-24-028-public-authorlist.tex}\end{sloppypar}
\end{document}

%% file: EXO-24-028-public-authorlist.tex
\cmsinstitute{Yerevan Physics Institute, Yerevan, Armenia}
{\tolerance=6000
A.~Hayrapetyan, V.~Makarenko\cmsorcid{0000-0002-8406-8605}, A.~Tumasyan\cmsAuthorMark{1}\cmsorcid{0009-0000-0684-6742}
\par}
\cmsinstitute{Institut f\"{u}r Hochenergiephysik, Vienna, Austria}
{\tolerance=6000
W.~Adam\cmsorcid{0000-0001-9099-4341}, L.~Benato\cmsorcid{0000-0001-5135-7489}, T.~Bergauer\cmsorcid{0000-0002-5786-0293}, M.~Dragicevic\cmsorcid{0000-0003-1967-6783}, P.S.~Hussain\cmsorcid{0000-0002-4825-5278}, M.~Jeitler\cmsAuthorMark{2}\cmsorcid{0000-0002-5141-9560}, N.~Krammer\cmsorcid{0000-0002-0548-0985}, A.~Li\cmsorcid{0000-0002-4547-116X}, D.~Liko\cmsorcid{0000-0002-3380-473X}, M.~Matthewman, J.~Schieck\cmsAuthorMark{2}\cmsorcid{0000-0002-1058-8093}, R.~Sch\"{o}fbeck\cmsAuthorMark{2}\cmsorcid{0000-0002-2332-8784}, M.~Shooshtari\cmsorcid{0009-0004-8882-4887}, M.~Sonawane\cmsorcid{0000-0003-0510-7010}, W.~Waltenberger\cmsorcid{0000-0002-6215-7228}, C.-E.~Wulz\cmsAuthorMark{2}\cmsorcid{0000-0001-9226-5812}
% ...
\par}
\vskip\cmsinstskip
\dag:~Deceased\\
$^{1}$Also at Yerevan State University, Yerevan, Armenia\\
$^{2}$Also at TU Wien, Vienna, Austria\\
% ...
